\documentclass[aps,prc,twocolumn,superscriptaddress,showpacs,floatfix,nofootinbib]{revtex4-1}
\usepackage{amsmath,graphicx,color}

\newcommand{\trento}{T$\mathrel{\protect\raisebox{-2.1pt}{R}}$ENTo}

\begin{document}

\title{Bayesian approach to long-range correlations and multiplicity fluctuations in nucleus-nucleus collisions}

\author{Kianusch Vahid Yousefnia}
\affiliation{Institut de physique th\'eorique, Universit\'e Paris Saclay, CNRS, CEA, F-91191 Gif-sur-Yvette, France}
\author{Atharva Kotibhaskar}
\affiliation{Department of Physics, Fergusson College, F.C. Road, Pune 411004, India}
\author{Rajeev Bhalerao}
\affiliation{Department of Physics, Indian Institute of Science Education and Research (IISER), Homi Bhabha Road, Pune 411008, India}
\author{Jean-Yves Ollitrault}
\affiliation{Institut de physique th\'eorique, Universit\'e Paris Saclay, CNRS, CEA, F-91191 Gif-sur-Yvette, France} 
\date{\today}

\begin{abstract}
  The number of particles detected in a nucleus-nucleus collision strongly depends on the impact parameter of the collision.
  Therefore, multiplicity fluctuations, as well as rapidity correlations of multiplicities, are dominated by impact parameter fluctuations.
  We present a method based on Bayesian inference which allows for a robust reconstruction of fluctuations and correlations at fixed impact parameter. 
  We apply the method to ATLAS data on the distribution of charged multiplicity and transverse energy. 
  We argue that multiplicity fluctuations are smaller at large rapidity than around central rapidity. 
  We suggest simple, new analyses, in order to confirm this effect. 
\end{abstract}
\maketitle

\section{Introduction}

The effective theory of strong interactions predicts that in ultrarelativistic nucleus-nucleus collisions, particle production occurs through the formation of color flux tubes parallel to the collision axis~\cite{Lappi:2006fp}, which are analogous to strings~\cite{Andersson:1983ia}. 
Since these tubes extend over a wide range in rapidity, one expects that particle production at different rapidities is strongly correlated~\cite{Gelis:2008sz,Lappi:2019kif}. 
The measurement of this rapidity-rapidity correlation~\cite{Bzdak:2012tp} has remained elusive so far because it is hidden by the trivial correlation induced by the variation of impact parameter $b$ within a given centrality class, since a more central collision produces more particles at all rapidities. 
Methods have been proposed to overcome this limitation~\cite{Olszewski:2017vyg}, but in practice, experimental results are essentially limited to the forward-backward correlation~\cite{PHOBOS:2006mfc,STAR:2009goo,ATLAS:2016rbh,ALICE:2017mtc}, which has reduced sensitivity to impact parameter fluctuations.

\begin{figure}[ht]
\begin{center}
\includegraphics[width=\linewidth]{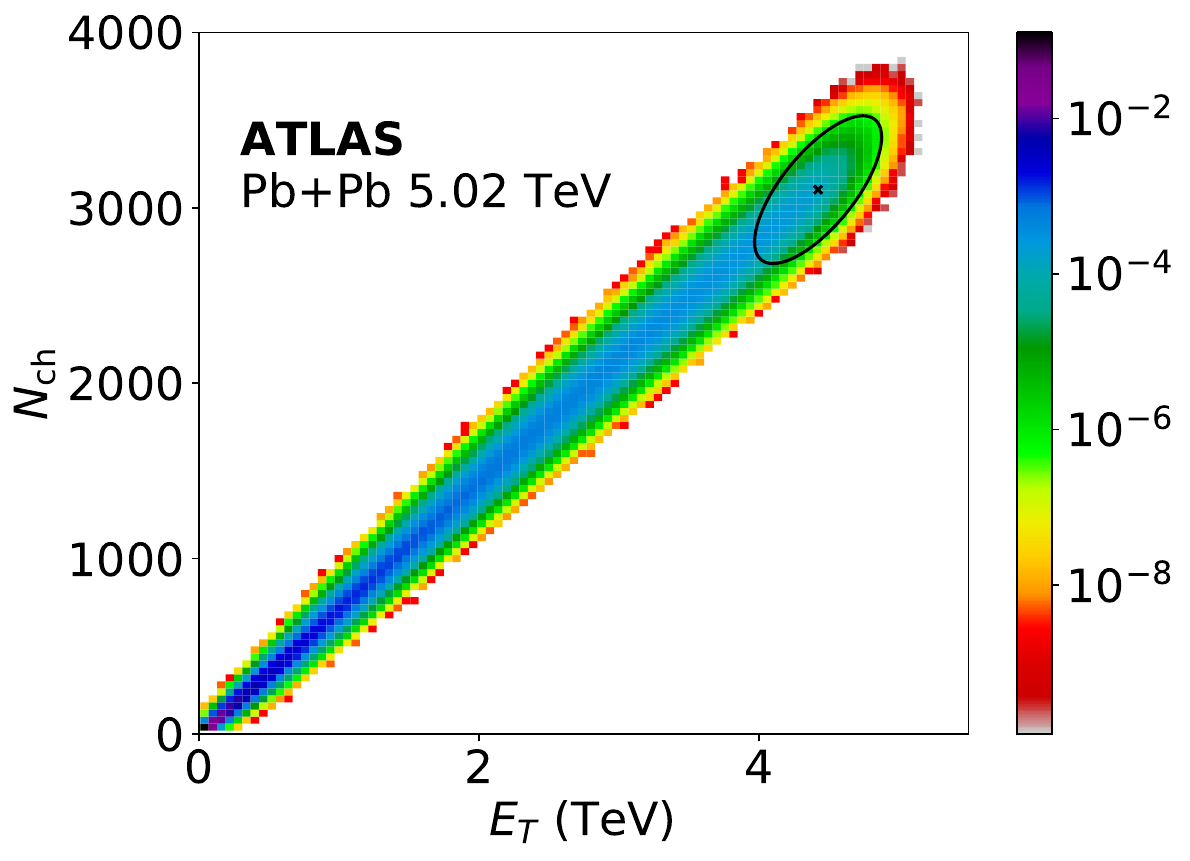} 
\end{center}
\caption{(Color online)
  Normalized histogram of the distribution of $(E_T,N_{\rm ch})$ measured by the ATLAS collaboration~\cite{Aaboud:2019sma} in Pb+Pb collisions at $\sqrt{s_{\rm NN}}=5.02$~TeV. $E_T$ denotes the transverse energy in the pseudorapidity window $3.2<|\eta|<4.9$, and $N_{\rm ch}$ the number of reconstructed tracks of charged particles in the pseudorapidity window $|\eta|<2.5$.
  The black ellipse is one of the results given by our Bayesian reconstruction. 
  It is the 99\% confidence ellipse for collisions at zero impact parameter. 
}
\label{fig:atlasillustration}
\end{figure}    
    
We introduce a simple Bayesian method which allows us to reconstruct correlations at fixed impact parameter, and therefore eliminates the trivial correlation due to variations of $b$. 
More specifically, if $(N_1,N_2,\cdots,N_p)$ denote multiplicities, or transverse energies, in $p$ rapidity windows, we show that their $p\times p$ covariance matrix can be reconstructed for $b=0$ (central collisions), and a $(p-1)\times (p-1)$ projection of this matrix for $b>0$. 
Note that Bayesian methods in the field of heavy-ion collisions usually involve sophisticated models and extensive calculations~\cite{Bernhard:2016tnd,Nijs:2020roc,JETSCAPE:2020mzn,Parkkila:2021tqq}. 
By contrast, the simple Bayesian approach implemented in this article does not involve any specific model of the collision dynamics. 
It could be implemented directly by experimental collaborations. 

We apply the reconstruction to ATLAS data~\cite{Aaboud:2019sma} on the distribution of the transverse energy $E_T$ measured in a calorimeter located at forward rapidity, and of the charged multiplicity $N_{\rm ch}$ measured in the central rapidity region.
The histogram of $(E_T,N_{\rm ch})$ for Pb+Pb collisions is represented in Fig.~\ref{fig:atlasillustration}. 
The impact parameter of a Pb+Pb collision is not measured, and this histogram is the superposition of the contributions of all impact parameters. 
However, we are able to reconstruct accurately the location of central collisions on this diagram, which is represented in the figure as a black curve corresponding to the 99\% confidence ellipse for collisions at $b=0$.
This is a specific example of what can be achieved through the Bayesian reconstruction. 

The method is explained in Sec.~\ref{s:method}. 
In Sec.~\ref{s:validation}, it is validated using a realistic model calculation in which the impact parameter is known, using the \trento{}~3D model of initial conditions~\cite{Ke:2016jrd}. 
In Sec.~\ref{s:data}, we apply it to ATLAS data. 
In Sec~\ref{s:correlations}, we present our results for the fluctuations of $E_T$ and $N_{\rm ch}$, 
and discuss what they tell us about the early collision dynamics. 
In Sec.~\ref{s:conclusions}, we suggest analyses which could be done easily with existing data, and which would shed new light on the rapidity dependence of fluctuations, and on long-range correlations.

\section{Method}
\label{s:method}

Detectors of heavy-ion experiments typically measure multiplicities of charged particles or energies deposited by these particles.
We assume that there are $p$ such observables in every event, which we denote by $(N_1,N_2,\cdots,N_p)$.
Experimentally, one typically measures the probability distribution of $(N_1,N_2,\cdots,N_p)$ in minimum-bias nucleus-nucleus collisions.
Our goal is to reconstruct as much as we can of the probability distribution of $(N_1,N_2,\cdots,N_p)$ at {\it fixed\/} impact parameter, without relying on any specific model of the collision dynamics.\footnote{Note that a quantity such as the number of participant nucleons $N_{\rm part}$~\cite{Miller:2007ri} or quarks~\cite{dEnterria:2020dwq} may be a better predictor of multiplicities than the impact parameter itself. Our analysis can be rephrased by replacing $b$ with a different variable such as $N_{\rm part}$. The price to pay is that the probability distribution of $N_{\rm part}$ is less simple than that of $b$, and depends on the details of the Glauber modeling.} 
The single-variable case, $p=1$, has been studied in~\cite{Das:2017ned}.
The present work generalizes this earlier study to several variables.
We illustrate this generalization on ATLAS data shown in Fig.~\ref{fig:atlasillustration}. 
These data correspond to the case $p=2$, where $N_1=E_T$ and $N_2=N_{\rm ch}$. 
Below, we will use either the general notation $(N_1,N_2)$, or the specific notation $(E_T,N_{\rm ch})$, depending on the context. 

The key assumption of our method is that for fixed impact parameter, fluctuations of $N_i$ are Gaussian. 
This can be viewed as a consequence of the central limit theorem.
In order for it to hold, we first need $N_i$ to be large enough. 
In the case of the ATLAS data represented in Fig.~\ref{fig:atlasillustration}, the multiplicity $N_{\rm ch}$ exceeds 3000 in central collisions, and an even larger number of particles contribute to the transverse energy $E_T$, so that this assumption is verified.
Even though the particles seen in detectors are emitted in the last stages of the collision dynamics~\cite{Busza:2018rrf}, there is a consensus that the origin of fluctuations lies in the early collision dynamics~\cite{Miller:2007ri,Broniowski:2007nz,Moreland:2014oya,Loizides:2017ack,Bierlich:2018xfw,Bozek:2019wyr}. 
Now, due to the strong Lorentz contraction at ultrarelativistic energies, the processes through which particle production occurs at various points in the transverse plane are causally disconnected and hence independent from each other.
Therefore, the multiplicity or transverse energy $N_i$ in some detector can be seen as the sum of a large number of independent contributions, which is the condition under which the central limit theorem applies.

More specifically, we assume that the probability distribution of $(N_1,N_2,\cdots,N_p)$ at fixed impact parameter is a multivariate normal distribution.
\begin{equation}
\label{nfixedb}
P(N_1,...,N_p|b)=
\frac{\exp\left(-\frac{1}{2}(N_i-\bar N_i(b))\Sigma^{-1}_{ij}(b)(N_j-\bar N_j(b))\right)}
{\sqrt{(2\pi)^{p} |\Sigma(b)| }},
\end{equation}
where, in the exponential, we use the Einstein summation convention over the repeated indices $i$ and $j$. 
In this equation, $\bar N_i$ is the mean, or average, value of $N_i$, and $\Sigma_{ij}$ is the symmetric covariance matrix:
\begin{eqnarray}
  \label{cov}
  \bar N_i&=&\langle N_i\rangle\cr
\Sigma_{ij}&=&\langle (N_i-\bar N_i)(N_j -\bar N_j)\rangle \cr
&=&\langle N_i N_j\rangle -\bar N_i \bar N_j,
\end{eqnarray}
where angular brackets denote an average over events with the same impact parameter $b$. 
$\Sigma^{-1}$ denotes the inverse matrix and $\left|\Sigma\right|$ the determinant. 
The validity of this Gaussian approximation will be discussed further in Sec.~\ref{s:validation}. 

The measured distribution is integrated over all values of impact parameter $b$. 
We carry out a simple change of variables and integrate instead over the cumulative probability distribution of $b$,
\begin{equation}
  \label{defcb}
  c_b\simeq \frac{\pi b^2}{\sigma_{\rm PbPb}},
\end{equation}
where $\sigma_{\rm PbPb}$ is the inelastic nucleus-nucleus cross section~\cite{Das:2017ned}.\footnote{Deviations from Eq.~(\ref{defcb}) only appear for very peripheral collisions which are excluded from our study.}
$c_b$ is referred to as the centrality fraction, or just centrality in the heavy-ion literature. 
The probability distribution of $c_b$ is uniform in the interval $[0,1]$. 
Therefore, the measured distribution is a simple integral over $c_b$: 
\begin{equation}
\label{cbint}
 P(N_1,\cdots,N_p)=\int_0^1 P(N_1,\cdots,N_p|c_b)dc_b.
\end{equation}
We further assume that the parameters in Eq.~(\ref{nfixedb}), namely, the mean values $\bar N_i$ and the elements of the covariance matrix $\Sigma_{ij}$, are smooth positive functions of $c_b$, which we choose to parametrize as the exponential of a polynomial~\cite{Das:2017ned}:
\begin{eqnarray}
  \label{exppoly}
\bar N_i(c_b)&=&\bar N_i(0)\exp\left(-\sum_{n=1}^{n_{\rm max}} a_{i,n} c_b^n\right)\cr
\Sigma_{ij}(c_b)&=&\Sigma_{ij}(0)\exp\left(-\sum_{m=1}^{m_{\rm max}} A_{i,j,m} c_b^m\right)
\end{eqnarray}
where $\bar N_i(0)$, $a_{i,n}$, $\Sigma_{ij}(0)$, $A_{i,j,m}$ are free parameters, and $n_{\rm max}$ and $m_{\rm max}$ are the degrees of the polynomials used to parametrize the mean and the covariance. 
The parameters are fitted in such a way that the distribution (\ref{cbint}) matches data, and the degree of the polynomial is adjusted so as to obtain a satisfactory fit.
We carry out a standard $\chi^2$ fit, keeping all non-empty boxes in the histogram of $(E_T,N_{\rm ch})$. 

As we shall see in Secs.~\ref{s:validation} and \ref{s:data}, the simple procedure defined by Eqs.~(\ref{nfixedb}), (\ref{cbint}) and (\ref{exppoly}) allows one to obtain a good fit to $P(N_1,\cdots,N_p)$.
The mean values $\bar N_i(c_b)$ returned by the fit closely match those obtained directly by  fixing the impact parameter and averaging $N_i$ over events, which implies that $\bar N_i(c_b)$ can be accurately reconstructed from data.
By contrast, one cannot reconstruct the impact parameter dependence of the whole covariance matrix $\Sigma_{ij}(c_b)$, and discrepancies are expected between the best fit value, and the value calculated directly by fixing the impact parameter. 
We conclude this section by explaining which information can be reconstructed, and which cannot. 

First, consider the tip of the distribution, corresponding to the largest values of $N_i$.
In this region, the integral in Eq.~(\ref{cbint}) is dominated by the contribution of central collisions, that is,  $c_b=0$.
Therefore, both $\bar N_i(0)$ and $\Sigma_{ij}(0)$ can be reconstructed using the tip of the distribution.
Looking at Fig.~\ref{fig:atlasillustration}, one clearly sees that the ellipse, which is defined by $\bar N_i(0)$ and $\Sigma_{ij}(0)$, closely fits the tip of the distribution of $(E_T,N_{\rm ch})$.  

\begin{figure}[ht]
\begin{center}
\includegraphics[width=\linewidth]{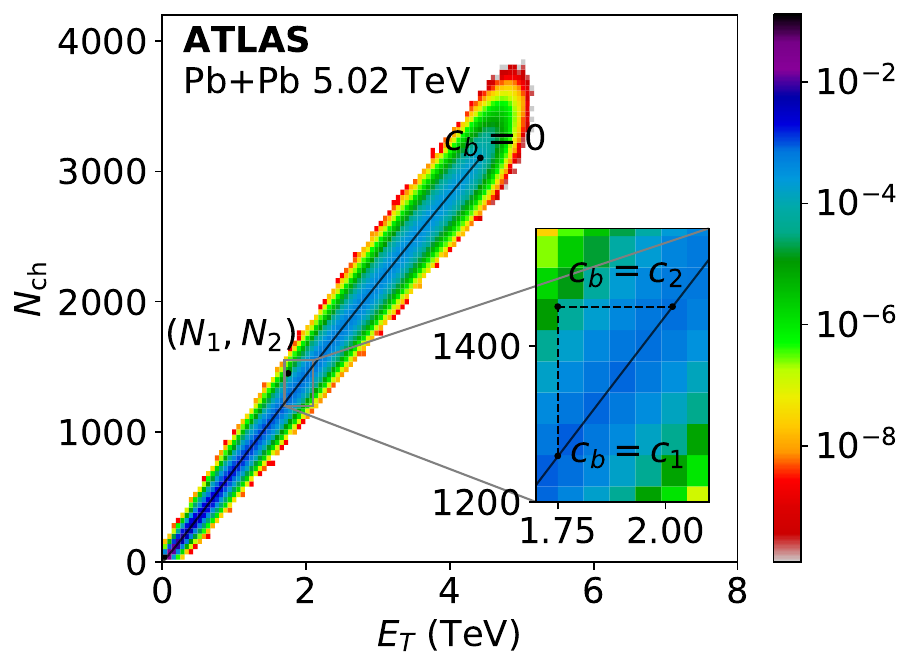} 
\end{center}
\caption{(Color online)
Illustration of Eq.~(\ref{defc1c2}). 
The line is the parametric curve $(\bar N_1(c_b),\bar N_2(c_b))$, or ridge line (see text). 
For a given point $(N_1,N_2)$ close to the ridge line, $c_1$ and $c_2$ defined by Eq.~(\ref{defc1c2}) are obtained by  projections on the ridge line.}
\label{fig:illusdeltac}
\end{figure}

Consider now the bulk of the distribution, corresponding to less central collisions. 
The mean value $\bar N_i(c_b)$ can be reconstructed from data in the following way. 
Define $c_i(N_i)$ as the cumulative probability distribution of $N_i$:
\begin{equation}
  \label{expcent}
  c_i(N_i)\equiv \int_{N_i}^{\infty} P(N_i)dN_i. 
\end{equation}
This is the usual definition of centrality in experiment~\cite{Abelev:2013qoq}. 
We use here the subscript $i$ because we consider the case where one measures several multiplicities $N_i$, each of which can be used as a centrality estimator $c_i$. 
Except for very peripheral and very central collisions, $c_i(N_i)$ defined by Eq.~(\ref{expcent}) is almost equal to the inverse function of $\bar N_i(c_b)$~\cite{Broniowski:2001ei,Das:2017ned}, that is; 
\begin{equation}
\label{defc1c2}
\bar N_i(c_i(N_i))\simeq N_i.
\end{equation}
Deviations are typically well below 1\%.
The functions $\bar N_i(c_b)$ define a ridge line which is displayed in Fig.~\ref{fig:illusdeltac}. 

By contrast, the centrality dependence of the covariance matrix cannot be fully reconstructed. 
This can be understood qualitatively as follows. 
A displacement parallel to the ridge line can be due either to a change in impact parameter, or to a fluctuation at fixed impact parameter. 
Therefore, only the projection of the covariance matrix orthogonal to the ridge line can be reconstructed.

We now explain this quantitatively in the case $p=2$, where each event is characterized by two multiplicities  $N_1$ and $N_2$. 
The centralities $c_1(N_1)$ and $c_2(N_2)$ defined by Eq.~(\ref{expcent}) are typically close to one another (see Fig.~\ref{fig:illusdeltac}). 
They coincide on the ridge line. 
The only sizable contribution to the integral in Eq.~(\ref{cbint}) is from values of $c_b$ close to $c_1(N_1)$ and $c_2(N_2)$.
We rewrite $N_i-\bar N_i(c_b)$ in Eq.~(\ref{nfixedb})  using  Eq.~(\ref{defc1c2}) and expand around $c_b$:  
\begin{eqnarray}
\label{expansion}
N_i-\bar N_i(c_b)&=&\bar N_i(c_i(N_i))-\bar N_i(c_b)  
\cr &\simeq &(c_i(N_i)-c_b) \bar N_i'(c_b), 
\end{eqnarray}
where $\bar N_i'$ is the derivative of $\bar N_i$ with respect to $c_b$ (note that $\bar N_i'<0$). 
Since the integrand in Eq.~(\ref{cbint}) is sizable only in a narrow range, one can neglect the variation of $\bar N_i'(c_b)$ and of $\Sigma_{ij}^{-1}(c_b)$ in this range.
Then, the integral over $c_b$ is a simple Gaussian integral, which is worked out in Appendix~\ref{s:saddlepoint}. 
In the case $p=2$, one obtains:
\begin{equation}
  \label{result2d}
P(N_1,N_2)\propto \exp\left(-\frac{\delta c^2}{2\sigma_\perp^2(c_b)}\right), 
\end{equation}
where $\delta c\equiv c_2(N_2)-c_1(N_1)$  represents the difference between the centralities defined by $E_T$ and $N_{\rm ch}$ in the case of the ATLAS analysis (see Fig.~\ref{fig:illusdeltac}), and 
\begin{equation}
  \label{defsigmaperp}
\sigma_\perp(c_b)\equiv\sqrt{ \frac{\Sigma_{11}(c_b)}{\bar N_1'(c_b)^2}+\frac{\Sigma_{22}(c_b)}{\bar N_2'(c_b)^2}-2\frac{\Sigma_{12}(c_b)}{\bar N_1'(c_b)\bar N_2'(c_b)}}
\end{equation}
represents the typical magnitude of $\delta c$. 
We now show that $\sigma_\perp(c_b)^2$ represents the projection on the covariance matrix perpendicular to the ridge line. 
The tangent to the ridge line is parallel to the vector ${\bf t}(c_b)\equiv\left(\bar N_1'(c_b),\bar N_2'(c_b)\right)$. 
The vector ${\bf n}(c_b)\equiv \left(1/\bar N_1'(c_b),-1/\bar N_2'(c_b)\right)$ is orthogonal to ${\bf t}(c_b)$.  
Now Eq.~(\ref{defsigmaperp}) can be rewritten as $\sigma_\perp(c_b)^2=n_i(c_b)\Sigma_{ij}(c_b)n_j(c_b)$, which is, up to a normalization, the projection of $\Sigma_{ij}(c_b)$ onto the direction of ${\bf n}(c_b)$, perpendicular to the ridge line

As we shall see in Sec.~\ref{s:validation}, the reconstruction of $\sigma_\perp(c_b)$ from the distribution of $E_T$ and $N_{\rm ch}$ is robust for all $c_b$, while the elements of the covariance matrix $\Sigma_{ij}(c_b)$ are well reconstructed only for $c_b=0$.

\section{Validation}
\label{s:validation}

\begin{figure}[ht]
\begin{center}
\includegraphics[width=\linewidth]{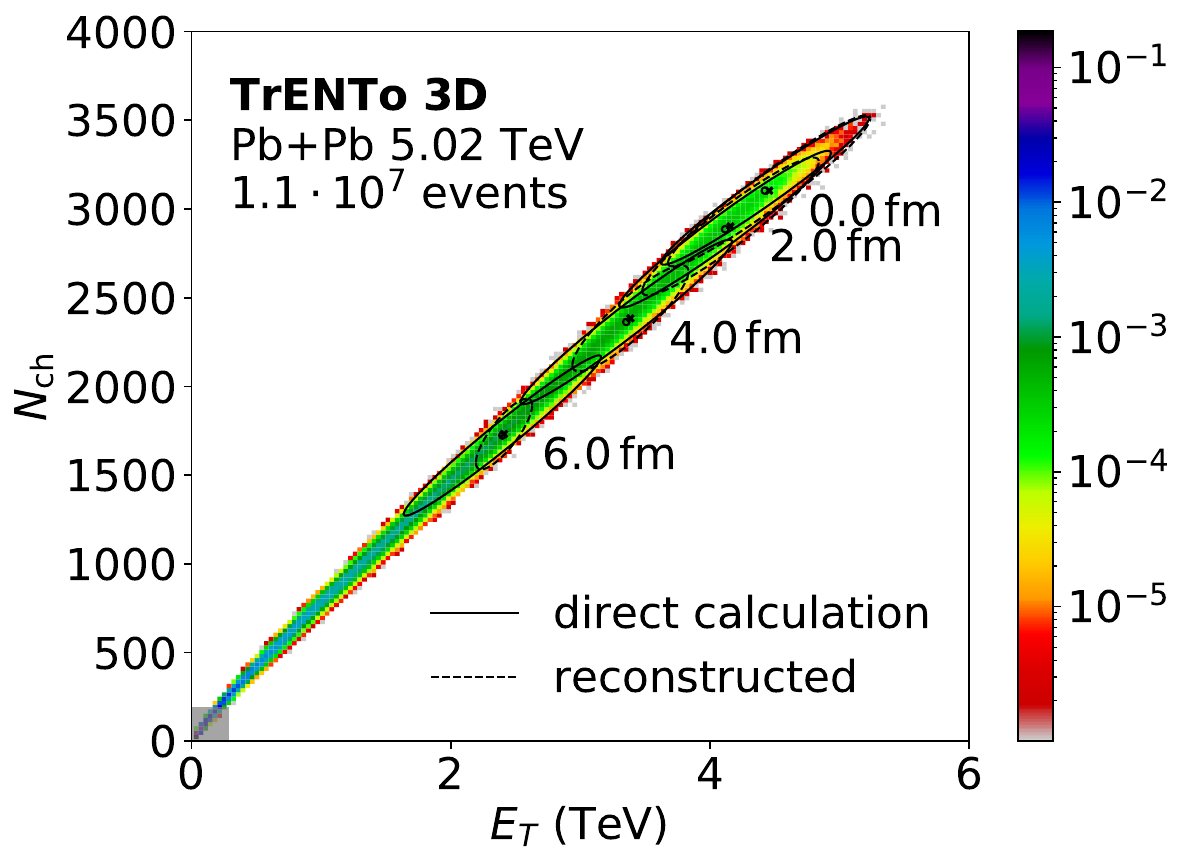} 
\end{center}
\caption{(Color online)
  Normalized histogram of $(E_T,N_{\rm ch})$ for $10^7$ minimum-bias Pb+Pb collisions at $\sqrt{s_{\rm NN}}=5.02$~TeV generated using the \trento{}~3D model~\cite{Ke:2016jrd}.
  Note that the number of events is smaller by a factor $\sim 16$ than in the data shown in Fig.~\ref{fig:atlasillustration}, and that we use a slightly finer binning in $(E_T,N_{\rm ch})$. 
   The grey square at the bottom left represents the part of the histogram which is excluded from the fit, corresponding to the most peripheral collisions.
  The ellipses correspond to the 99\% confidence ellipses for fixed values of $b=0,2,4,6$~fm, calculated directly (full lines, circles at the center) or from the Bayesian reconstruction (dashed lines, crosses at the center). 
}
\label{fig:histogramtrento3d}
\end{figure}    

In order to validate the reconstruction outlined in Sec.~\ref{s:method}, we test it using fake data generated using the \trento{}~3D model~\cite{Ke:2016jrd}.
This is a state of the art Monte Carlo generator of the initial state of proton-nucleus and nucleus-nucleus collisions, which has been tuned to reproduce several rapidity-dependent observables. 
It returns, for each collision event, an entropy density profile $s(x,y,\eta)$ in three dimensions at an early time after the collision, where $(x,y)$ are cartesian coordinates in the transverse plane and $\eta$ is the space-time rapidity. 
We generate $10^7$ minimum-bias Pb+Pb collisions at $\sqrt{s_{\rm NN}}=5.02$~TeV, and we convert the model predictions into values of $E_T$ and $N_{\rm ch}$, that can be compared with the ATLAS data in Fig.~\ref{fig:atlasillustration}  in the following way.
We assume that $N_{\rm ch}$ is proportional to the initial entropy, obtained by integrating $s(x,y,\eta)$ over $(x,y)$ and over the interval $|\eta|<2.5$.
We assume that $E_T$ is proportional to the initial energy, obtained by integrating $s(x,y,\eta)^{4/3}$ over $(x,y)$ and over the interval $3.2<|\eta|<4.9$.\footnote{We thereby assume that the equation of state is conformal, which is approximately true at the high temperatures achieved in the early stages of the collision.} 
Finally,  we normalize $E_T$ and $N_{\rm ch}$ in such a way that their mean values for central collisions coincide with those reconstructed from ATLAS data (Sec.~\ref{s:data}). 
The resulting histogram of $(E_T,N_{\rm ch})$ is displayed in Fig.~\ref{fig:histogramtrento3d}.  
It is roughly similar to the experimental distribution in  Fig.~\ref{fig:atlasillustration}.  

\begin{figure}[ht]
\begin{center}
\includegraphics[width=\linewidth]{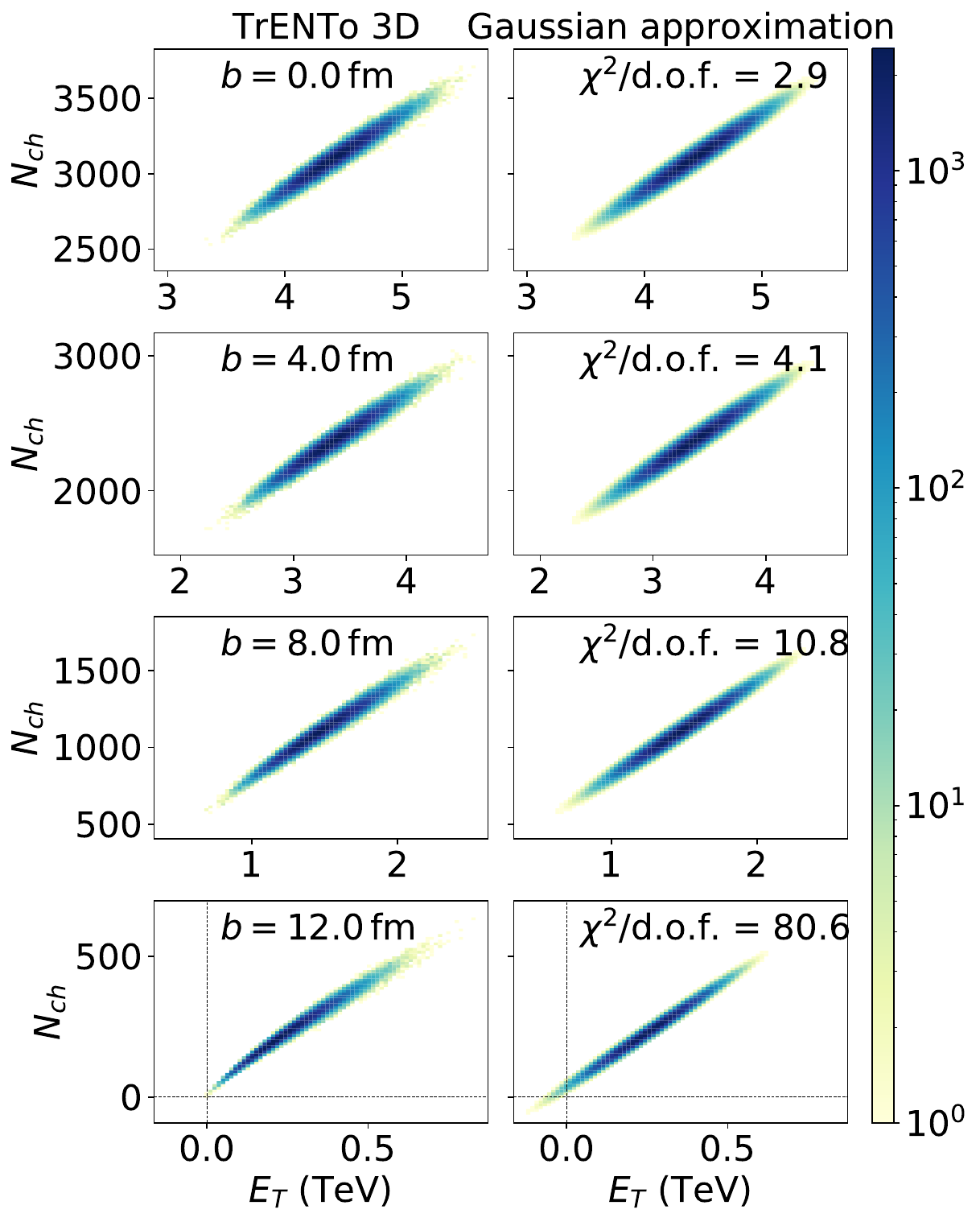} 
\end{center}
\caption{(Color online)
Left: Histogram of the distribution of $(E_T,N_{\rm ch})$ from the \trento{}~3D model of initial conditions~\cite{Ke:2016jrd} (see Sec.~\ref{s:validation} for details) in Pb+Pb collisions at $\sqrt{s_{\rm NN}}=5.02$~TeV for fixed values of the impact parameter, from top to bottom: $b=0,4,8,12$~fm. Right: two-dimensional Gaussian fits to these distributions. 
We indicate the $\chi^2$ of each fit. 
}
\label{fig:gaussiantest}
\end{figure}    

In order to check the validity of the assumption of Gaussian fluctuations, Eq.~(\ref{nfixedb}),  we also run the model for fixed impact parameter. 
We have generated $10^5$ events for each value of $b$. 
The right panels of Fig.~\ref{fig:gaussiantest} display Gaussian fits to the left panels.
The quality of the fits, measured by their $\chi^2$, decreases as the impact parameter increases.
The Gaussian approximation captures the distribution except for the bottom panel, corresponding to $b=12$~fm, where deviations are sizable.
For this value of $b$, the Gaussian extends to negative values of $E_T$ and $N_{\rm ch}$ which are unphysical.\footnote{In the single-variable case this could in principle be fixed by replacing the Gaussian with a gamma distribution~\cite{Rogly:2018ddx}, but we have not found a simple way of generalizing this approach to the multi-variable case.}
Therefore, the assumption of Gaussian fluctuations breaks down for peripheral collisions, which we exclude from the fit.  
Excluded events are represented as a grey square at the bottom left of Fig.~\ref{fig:histogramtrento3d}, corresponding to events with $E_T<E_{T,{\rm min}}=0.289$~TeV and $N_{\rm ch}<N_{\rm ch,min}=194.39$. 

We then fit the remaining distribution using the method outlined in Sec.~\ref{s:method}.
We choose $n_{\rm max}=3$ and $m_{\rm max}=2$ in Eq.~(\ref{exppoly}).
Therefore, there are 4 parameters for $\bar E_T(c_b)$ and $\bar N_{\rm ch}(c_b)$, and 3 parameters for each element of the covariance matrix, which gives a total of 17 fit parameters. 
Note that we need to evaluate the Pb+Pb cross section in order to convert the impact parameter into a centrality fraction using Eq.~\ref{defcb}.
This is done by computing the fraction $f(b)$ of events whose impact parameter is below some value $b$, small enough that the probability of collision is very close to unity. 
The cross section is then evaluated as $\sigma_{\rm PbPb}=\pi b^2/f(b)$ and we obtain $\sigma_\text{PbPb}=(800\pm 2)\,\text{fm}^2$.

\begin{figure*}[ht]
\begin{center}
\includegraphics[width=\linewidth]{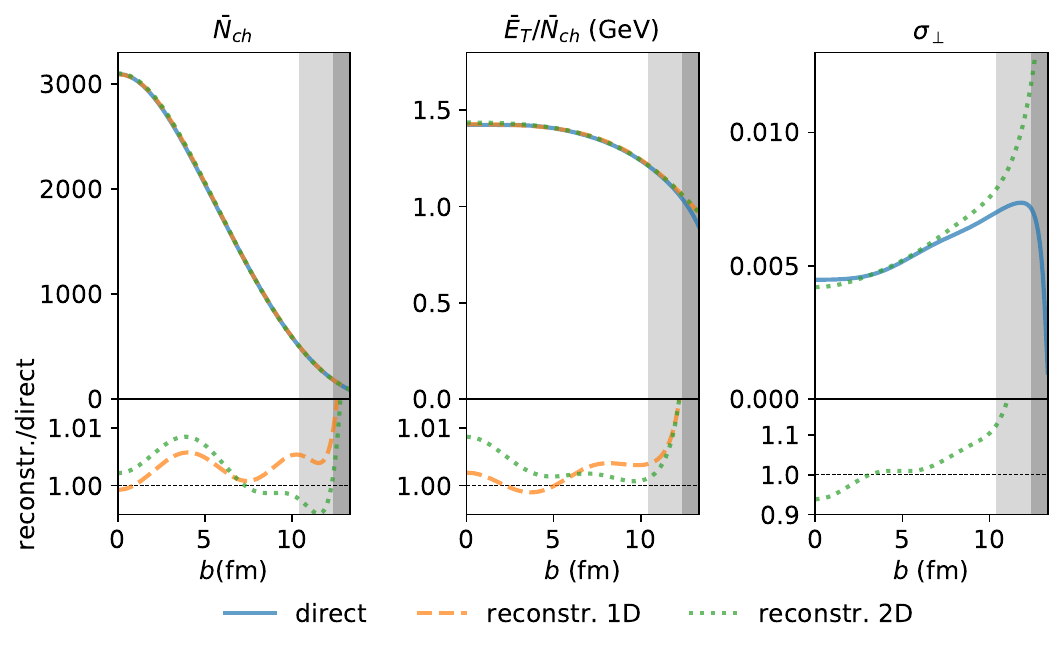} 
\end{center}
\caption{(Color online)
  Left to right: $\bar N_{\rm ch}(b)$, $\bar E_T(b)/\bar N_{\rm ch}(b)$ and $\sigma_\perp(b)$ (defined by Eq.~(\ref{defsigmaperp})) versus impact parameter.
The vertical grey bands indicate the limits above which the reconstruction can no longer be trusted because peripheral events have been excluded from the fit (see text for details). 
  The full lines correspond to the direct calculation, where one averages over events at fixed $b$.
  The derivatives $\bar N_{\rm ch}'(c_b)$ and $\bar E_T'(c_b)$ in Eq.~(\ref{defsigmaperp}) are evaluated by fitting $\bar N_{\rm ch}(c_b)$ and $\bar E_T(c_b)$ with a smooth curve and taking the derivative of the fit.
  The dashed lines in the left and middle panels correspond to the values reconstructed using the projected distributions $P(E_T)$ and $P(N_{\rm ch})$, following the same method as in Ref.~\cite{Das:2017ned}.
  The dotted lines correspond to the values reconstructed using the distribution $P(E_T,N_{\rm ch})$, which is done as explained in Sec.~\ref{s:method}.
  The bottom panels display the ratio between the reconstructed value and the direct calculation. 
}
\label{fig:directvsreconstructed}
\end{figure*}    

In order to assess the quality of the reconstruction, we first evaluate the mean values of $E_T$ and $N_{\rm ch}$ and their covariance matrix using Eq.~(\ref{cov}), where one fixes the impact parameter before averaging over events.
We compare these results with those obtained by fitting the distribution as explained in Sec.~\ref{s:method}. 
Since we have excluded peripheral events, we do not expect the reconstruction to be valid if $b$ is too large.
However, the exclusion of peripheral events is done according to the values of $E_T$ and $N_{\rm ch}$, not to the value of impact parameter.
We expect that the reconstruction is robust for values of $b$ such that at least 90\% of events are included. 
The limiting value of $b$, above which this condition is no longer satisfied, is represented as light grey bands in Fig.~\ref{fig:directvsreconstructed}.
The dark grey band is defined by $\bar E_{T}(b)<E_{T,\rm min}$.  

\begin{figure}[ht]
\begin{center}
\includegraphics[width=\linewidth]{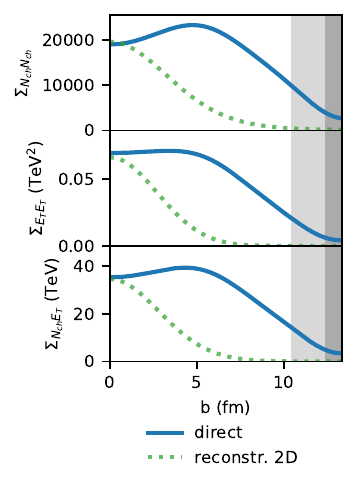} 
\end{center}
\caption{(Color online)
  Elements of the covariance matrix as a function of impact parameter in the \trento{}~3D model.
  Top to bottom: Variance of $N_{\rm ch}$, variance of $E_T$, covariance of $N_{\rm ch}$ and $E_T$.
  The full lines correspond to the direct calculation, where one averages over events at fixed $b$.
  The dotted lines correspond to the reconstructed values. 
}
\label{fig:covariancestrento3d}
\end{figure}    

Figure~\ref{fig:directvsreconstructed} shows that the mean values $\bar N_{\rm ch}(b)$ and $\bar E_T(b)$ are accurately reconstructed (we  plot the ratio $\bar E_T(b)/\bar N_{\rm ch}(b)$ rather than $\bar E_T(b)$ for reasons that will appear in Sec.~\ref{s:data}), as well as the width $\sigma_\perp(b)$ defined by Eq.~(\ref{defsigmaperp}). 
We have argued on general grounds that one cannot expect to reconstruct the elements of the covariance matrix $\Sigma_{ij}$, except for central collisions.
This is confirmed by the results displayed in Fig.~\ref{fig:covariancestrento3d}.
All three elements of the covariance matrix are reconstructed with an error smaller than 6\% for $b=0$.
However, as $b$ increases, the reconstructed values quickly deviate from the direct calculation.
This can also be seen by looking at the ellipses in Fig.~\ref{fig:histogramtrento3d}. 
They are the 99\% confidence ellipses (that is, ellipses that contain 99\% of the events) for several values of $b$.  
They are defined by 
\begin{equation}
\label{defconfidenceellipse}
(N_i-\bar N_i(b))\Sigma^{-1}_{ij}(b)(N_j-\bar N_j(b)) =-2 \ln(1-0.99).
\end{equation}
For $b=0$, the reconstructed ellipse almost coincides with the direct calculation, but the discrepancy quickly increases as $b$ increases. 

\section{Application to ATLAS data} 
\label{s:data}

\begin{figure}[ht]
\begin{center}
\includegraphics[width=\linewidth]{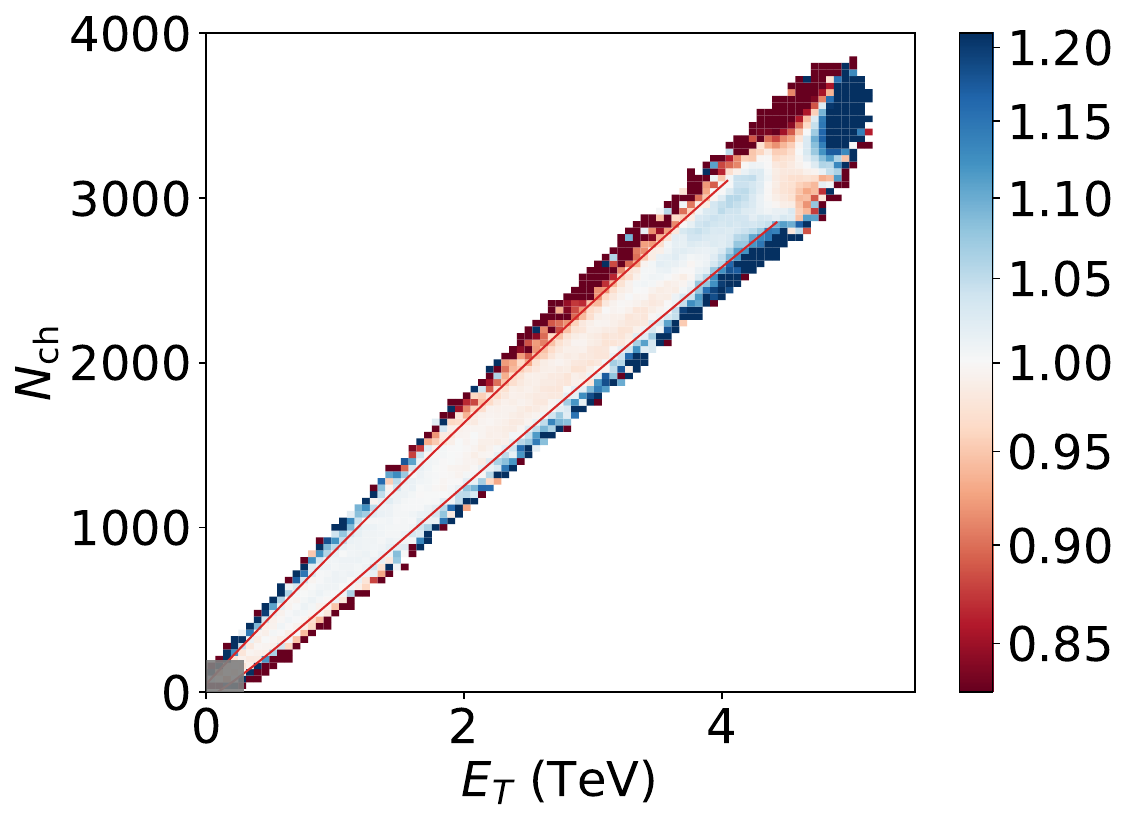} 
\end{center}
\caption{(Color online)
  Ratio of the fit to ATLAS data in Fig.~\ref{fig:atlasillustration}.
  As in Fig.~\ref{fig:histogramtrento3d}, the grey square at the bottom left represents the part of the histogram which is excluded from the fit, corresponding to the most peripheral collisions.
  The lines correspond to contours with $\delta c=2.6\sigma_\perp$ in Eq.~(\ref{result2d}), so that the region between the lines contains $\sim 99\%$ of the events.
  Specifically, the upper and lower lines are the parametric curves $\left(\bar E_T(c_b)+2.6\sigma_\perp(c_b)\bar E_T'(c_b),\bar N_{\rm ch}(c_b)\right)$, and $\left(\bar E_T(c_b), \bar N_{\rm ch}(c_b)+2.6\sigma_\perp(c_b)\bar N_{\rm ch}'(c_b)\right)$. 
}
\label{fig:fitoverdata}
\end{figure}    
We finally apply the Bayesian reconstruction to the ATLAS data shown in Fig.~\ref{fig:atlasillustration}.
Some of the low multiplicity events correspond to photo-nuclear events which do not contribute to the inelastic cross section. 
Therefore, we evaluate the total number of inelastic collisions $N_{\rm eve}$ using the same centrality calibration as the ATLAS collaboration, which gives $N_{\rm eve}=158,568,641$. 
We exclude peripheral events using the same cuts as in the \trento{}~3D calculation.
Roughly 40\% of the events are excluded. 
Figure~\ref{fig:fitoverdata} displays the ratio of the fit to data.
The fit is excellent, except far out in the tail of the distribution. 
The deviation between fit and data is at the percent level between the two lines, which are drawn is such a way that they encompass 99\% of the events. 

\begin{table}
\begin{tabular}{|c|c|c|}
\hline
&ATLAS&\trento{} 3D\cr
\hline
$\bar E_T(0)$& 4.424(33) TeV &  4.457(33) TeV \cr
$\bar N_{\rm ch}(0)$& 3104(1) & 3103(1) \cr
$\Sigma_{E_TE_T}(0)$&$0.0224(11)$~TeV$^2$&$0.0663(34)$ ~TeV$^2$\cr
$\Sigma_{N_{\rm ch}N_{\rm ch}}(0)$&$19236(759)$&$19664(776)$ \cr
$\Sigma_{E_TN_{\rm ch}}(0)$&$14.71(28)$~TeV&$ 34.62(70)$~TeV \cr
$a_{E_T,1}$&$4.00$&$4.43$ \cr
$a_{N_{\rm ch},1}$&$3.77$& $4.29$\cr
$a_{E_T,2}$&$-1.49$&$-1.19$\cr
$a_{N_{\rm ch},2}$&$-0.91$&$-1.78$  \cr
$a_{E_T,3}$&$3.97$&$4.17$ \cr
$a_{N_{\rm ch},3}$&$3.86$&$4.15$ \cr
$A_{E_T,E_T,1}$&$4.08$& $17.33$\cr
$A_{N_{\rm ch},N_{\rm ch},1}$&$7.18$& $11.71$\cr
$A_{E_T,N_{\rm ch},1}$&$7.48$&  $15.94$\cr
$A_{E_T,E_T,2}$&$1.18$& $5.55$ \cr
$A_{N_{\rm ch},N_{\rm ch},2}$&$-1.89$&$-6.90$ \cr
$A_{E_T,N_{\rm ch},2}$&$1.32$&$-2.14$ \cr
\hline
\end{tabular}
\caption{\label{fitparameters} 
  Values of fit parameters, as defined by Eq.~(\ref{exppoly}), for ATLAS data shown in Fig.~\ref{fig:atlasillustration}, and 
 for the \trento{}~3D calculation shown in Fig.~\ref{fig:histogramtrento3d}.  The uncertainties on the first five parameters are estimated using the relative difference between the direct calculation and its reconstruction in the \trento{}~3D model. We omitted error bars on the polynomial coefficients in Eq.~(\ref{exppoly}) because the individual uncertainties are not reflective of the uncertainty of the corresponding physical observable, e.g. $\bar{E}_T$.
}
\end{table}

The values of the fit parameters are listed in Table~\ref{fitparameters}. 
As explained in Sec.~\ref{s:method}, the output of the fit consists of the following information: 
\begin{itemize}
\item Impact parameter dependence of the mean transverse energy $\bar E_T$ and of the mean charged multiplicity $\bar N_{\rm ch}$. 
\item Variances of $E_T$ and $N_{\rm ch}$, and covariance of $E_T$ and $N_{\rm ch}$, for central collisions at $b=0$. 
\item Impact parameter dependence of the width $\sigma_\perp$ defined by Eq.~(\ref{defsigmaperp}).
\end{itemize}
We estimate the error on these quantities as follows. 
First, we check the robustness of the fit by varying the cutoffs in $E_T$ and $N_{\rm ch}$, and we exclude a fraction of events varying from   35\% to 45\% of the events. 
The variation of the results is significantly smaller than the difference between the direct calculation and the reconstruction in the \trento{}~3D calculation.
Therefore, we estimate the relative error on our result as the maximum relative error between the direct calculation and the reconstructed value in the \trento{}~3D calculation, shown in Figs.~\ref{fig:directvsreconstructed} and \ref{fig:covariancestrento3d}.

In the remainder of this section, we discuss our results for the mean values of  $N_{\rm ch}$ and $E_T$. 
Note that these results could also have been obtained directly from the one-dimensional projections of the two dimensional histogram in Fig.~\ref{fig:atlasillustration}, i.e., from the histograms of $N_{\rm ch}$ and $E_T$ alone~\cite{Das:2017ned}. 
The most important results, which involve the covariance matrix of $N_{\rm ch}$ and $E_T$, will be discussed in Sec.~\ref{s:correlations}.

\begin{figure}[ht]
\begin{center}
\includegraphics[width=\linewidth]{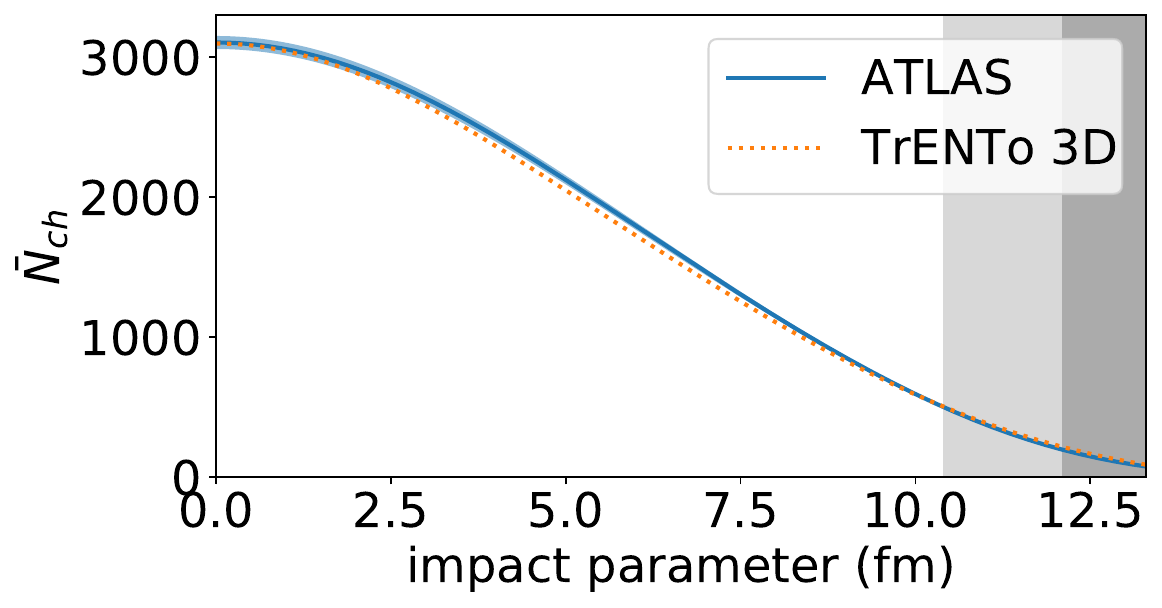} 
\end{center}
\caption{(Color online)
  Variation of the mean charged multiplicity as a function of impact parameter.
  The full line is the value reconstructed using ATLAS data. The shaded band is our estimate of the error band (see text).
  The dotted line is the result of the direct \trento{}~3D calculation, shown as a full line in Fig.~\ref{fig:directvsreconstructed}. 
}
\label{fig:nchatlas}
\end{figure}  

\begin{figure}[ht]
\begin{center}
\includegraphics[width=\linewidth]{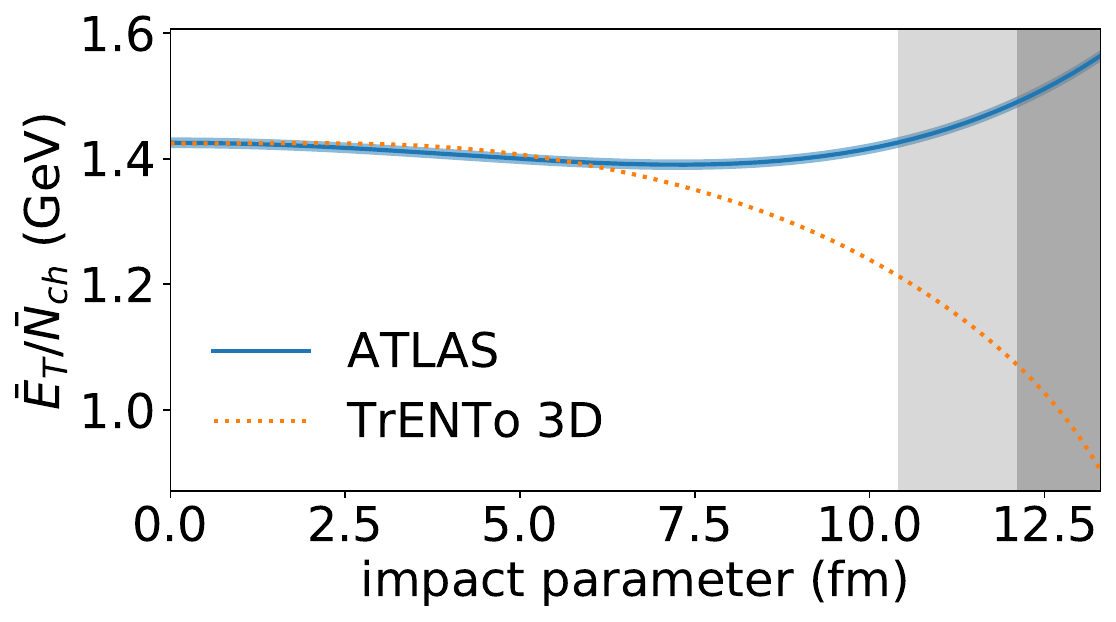} 
\end{center}
\caption{(Color online)
  Same as Fig.~\ref{fig:nchatlas} for the ratio of the mean transverse energy to the mean charged multiplicity. 
}
\label{fig:etnchatlas}
\end{figure}  

Figure~\ref{fig:nchatlas} displays the mean charged multiplicity as a function of impact parameter. 
For the conversion between centrality fraction and impact parameter, we have used the value of the inelastic Pb+Pb cross section $\sigma_{\rm PbPb}=767$~fm$^2$ extracted from a Glauber calculation~\cite{ALICE:2015juo}.\footnote{Note that this value of $\sigma_{\rm PbPb}$ differs from that returned by the \trento{}~3D calculation. We compare model and experiment at the same impact parameter, not at the same centrality fraction.} 
Note that $N_{\rm ch}$ counts reconstructed tracks, and that the efficiency of the reconstruction quickly decreases for $p_t<0.8$~GeV/c~\cite{Aaboud:2019sma}. 
Therefore, the value of $N_{\rm ch}$ seen by ATLAS is only $\sim 29\%$ of that seen by ALICE~\cite{ALICE:2016fbt} in the corresponding pseudorapidity interval. 
For the sake of comparison, we also show the results from the \trento{}~3D calculation (corresponding to the full lines in Fig.~\ref{fig:directvsreconstructed}).
The variation of $\bar N_{\rm ch}$ as a function of $b$ is well reproduced by the model, although  not perfectly. 

Figure~\ref{fig:etnchatlas} displays the ratio of the mean transverse energy to the mean charged multiplicity, as a function of impact parameter. 
Let us first comment on the order of magnitude of the result for $b=0$. 
The ratio $\bar E_T/\bar N_{\rm ch}$ can be decomposed as $(\bar E_T/\bar N_{\rm FCAL})(\bar N_{\rm FCAL}/\bar N_{\rm ch})$, where $N_{\rm FCAL}$ is the hadron multiplicity (neutral and charged) falling into the calorimeter acceptance.
We evaluate $N_{\rm FCAL}$ by integrating the pseudorapidity spectra of charged particles~\cite{ALICE:2016fbt} over the acceptance covered by the calorimeter, multiplying by a factor $\frac{3}{2}$ to take into account neutral particles. 
We then evaluate its average value at $b=0$ by extrapolating linearly the values in the centrality intervals $0-5\%$ and $5-10\%$, that is, $\bar N_{\rm FCAL}(0)\simeq \frac{3}{2}N_{\rm FCAL}(0-5\%)-\frac{1}{2}N_{\rm FCAL}(5-10\%)$. 
We obtain $\bar N_{\rm FCAL}(b=0)\simeq 7930$. 
The contribution of a hadron to the transverse energy is roughly the transverse mass $m_t=\sqrt{p_t^2+m^2}$. 
Using the value of $\bar E_T(0)$ in Table~\ref{fitparameters}, we obtain $\bar E_T(0)/\bar N_{\rm FCAL}(0)=0.558$~GeV, which is the expected order of magnitude for the average transverse mass. 

We next comment on the dependence of the ratio $\bar E_T/\bar N_{\rm ch}$ on impact parameter.
ATLAS data show a non-monotonic behavior, which is not reproduced by the \trento{}~3D model.  
This variation can be understood by decomposing again $E_T/N_{\rm ch}$ as $(E_T/N_{\rm FCAL})(N_{\rm FCAL}/N_{\rm ch})$. 
The ratio $E_T/N_{\rm FCAL}$ is the average transverse mass,  which is determined by the mean transverse momentum $\langle p_t\rangle$, which itself decreases mildly as a function of impact parameter~\cite{ALICE:2018hza}. 
This effect is responsible for the decrease seen in the model, and in data at small $b$.
The ratio $N_{\rm FCAL}/N_{\rm ch}$ is the ratio of the multiplicity in the forward rapidity region, covered by the calorimeter, and the central rapidity region,  where $N_{\rm ch}$ is measured. 
Now, the rapidity distribution becomes slightly broader as the impact parameter increases~\cite{ALICE:2016fbt}. 
The intuitive picture for this phenomenon is that the stopping between the nuclei is not as strong as in central collisions. 
This implies that the ratio  $N_{\rm FCAL}/N_{\rm ch}$ increases as a function of impact parameter. 
This effect is not reproduced by the \trento{}~3D model. 
It overrides the decrease of $\langle p_t\rangle$ for large values of $b$, leading to the increase seen in ATLAS data. 

\section{Fluctuations at large and central rapidities}
\label{s:correlations}

In this Section, we present results involving the covariance matrix of $E_T$ and $N_{\rm ch}$. 
These are our most important results, as they shed light on the fluctuations in the central rapidity region probed by $N_{\rm ch}$, in the large rapidity region probed by  $E_T$, and on their mutual correlations. 
We first isolate dynamical fluctuations of $E_T$ and  $N_{\rm ch}$ by subtracting out Poisson fluctuations in Sec.~\ref{s:subtraction}. 
We then present our results for central collisions in Sec.~\ref{s:b=0}, and finally the results on the impact parameter dependence in Sec.~\ref{s:sigmaperp}. 

\subsection{Subtraction of Poisson fluctuations}
\label{s:subtraction}

We generally consider quantities of the form $N=\sum_{i=1}^M x_i$, where $M$ is the particle multiplicity. 
Our goal is to identify non-trivial fluctuations and correlations of these quantities. 
The baseline is the case where the probabilities of finding a particle in different regions of phase space are independent variables~\cite{DiFrancesco:2016srj}. 
In this case, $x_i$ are independent variables and $M$ follows a Poisson distribution.
A simple calculation shows that the variance of $N$ is the expectation value of $\sum_{i=1}^M x_i^2$. 
We refer to this contribution loosely as  ``the variance of Poisson fluctuations'', and we subtract it from the observed variance in order to isolate the non-trivial part. 

For the charged multiplicity, one simply counts particles, so that $x_i=1$, and the subtraction is straightforward:
\begin{equation}
\label{nchsubtraction}
\Sigma_{N_{\rm ch}N_{\rm ch}}(b)\rightarrow\Sigma_{N_{\rm ch}N_{\rm ch}}(b)-\bar N_{\rm ch}(b).
\end{equation}
There is however an uncertainty on this subtraction due to hadronic decays. 
If a hadron decays into two charged particles, which both fall into the detector acceptance, then the contribution of that hadron is $x_i=2$, not $x_i=1$.
The magnitude of this effect is modest, and cannot be evaluated accurately, because it depends on the hadronization mechanism. 
In the case of a fluid-dynamical model, it depends on the freeze-out temperature~\cite{Mazeliauskas:2018irt}. 
We include this uncertainty in our error bar, by multiplying $\bar N_{\rm ch}(b)$ in Eq.~(\ref{nchsubtraction}) by a coefficient which can vary between $1$ and $1.2$. 

The transverse energy can be written as $E_T=\sum_i x_i$, where the sum has $N_{\rm FCAL}$ terms, corresponding to all  hadrons, neutral and charged, falling in the calorimeter, and $x_i\simeq m_t$, where $m_t=\sqrt{p_t^2+m^2}$ is the transverse mass.  
The average value of $E_T$ at fixed $b$ is $\bar E_T(b)=\langle m_t\rangle \bar N_{\rm FCAL}(b)$, and the variance of Poisson fluctuations is  $\langle m_t^2\rangle \bar N_{\rm FCAL}(b)$, where angular brackets denote an average value over hadrons in the calorimeter. 
This quantity can be decomposed as 
\begin{equation}
\label{decomp}
\langle m_t^2\rangle \bar N_{\rm FCAL}(b)=\frac{\langle m_t^2\rangle}{\langle m_t\rangle^2}\frac{\bar N_{\rm ch}(b)}{\bar N_{\rm FCAL}(b)}\frac{\bar E_T(b)^2}{\bar N_{\rm ch}(b)}.
\end{equation}
We evaluate the first factor on the right-hand side using ALICE data on identified particle spectra~\cite{ALICE:2013mez}, which give $\langle m_t^2\rangle/\langle m_t\rangle^2\simeq 1.52$ for central collisions at $\sqrt{s_{\rm NN}}=2.76$~TeV. We neglect the dependence of this ratio on impact parameter, rapidity and $\sqrt{s_{\rm NN}}$. 
The second factor can be estimated for central collisions using the value of $\bar N_{\rm ch}(0)$ in Table~\ref{fitparameters}, and the estimate $\bar N_{\rm FCAL}(0)\simeq 7930$ obtained in Sec.~\ref{s:data}. 
We also neglect its dependence on rapidity and impact parameter. 
Putting these factors together, the subtraction of Poisson fluctuations is done according to the formula: 
\begin{equation}
\label{etsubtraction}
\Sigma_{E_TE_T}(b)\rightarrow\Sigma_{E_TE_T}(b)-\alpha\frac{\bar E_T(b)^2}{\bar N_{\rm ch}(b)}, 
\end{equation}
where $\alpha\simeq 0.59$. We assign an uncertainty of $\pm 20\%$ to this factor $\alpha$. 

Finally, since $E_T$ and $N_{\rm ch}$ are measured in separate rapidity regions, no hadron contributes simultaneously to  $E_T$ and $N_{\rm ch}$, and Poisson fluctuations do not contribute to the covariance $\Sigma_{E_TN_{\rm ch}}(b)$. 

Using the values in Table~\ref{fitparameters}, for central collisions, the subtracted quantities in Eqs.~(\ref{nchsubtraction}) and (\ref{etsubtraction}) are $16\%$ and $17\%$ of the total. 
This implies that at least $83\%$ of the variances of $E_T$ and $N_{\rm ch}$ can be attributed to dynamical fluctuations. 

\subsection{Fluctuations of $N_{\rm ch}$ and $E_T$ in central collisions}
\label{s:b=0}

\begin{table}[ht]
\begin{tabular}{|c||c|c|}
\hline
&$E_T$&$N_{\rm ch}$\cr
\hline
\hline
&$1.14(6)\times 10^{-3}$&$1.07(2)\times 10^{-3}$\cr
$E_T$&$\mathbf{0.95(7)\times 10^{-3}}$&$\mathbf{1.07(2)\times 10^{-3}}$\cr
&$\mathit{(3.57\times 10^{-3})}$&$\mathit{(2.52\times 10^{-3})}$\cr
\hline
&$1.07(2)\times 10^{-3}$&$2.00(8)\times 10^{-3}$\cr
$N_{\rm ch}$&$\mathbf{1.07(2)\times 10^{-3}}$&$\mathbf{1.68(9)\times 10^{-3}}$\cr
&$\mathit{(2.52\times 10^{-3})}$&$\mathit{(1.96\times 10^{-3})}$ \cr
\hline
\end{tabular}
\caption{\label{covariance} 
Relative covariance matrix $\sigma_{ij}$ of $E_T$ and $N_{\rm ch}$, defined by Eq.~(\ref{defsigmaij}), for central Pb+Pb collisions ($b=0$). 
For each element, the first line is the total covariance returned by the fit to data. 
The second line is the dynamical covariance obtained after subtracting Poisson fluctuations (see Sec.~\ref{s:subtraction}). 
The third line is the value obtained in the \trento{}~3D calculation by simulating events at $b=0$, which differs by a few percent from the reconstructed value in Table~\ref{fitparameters}. 
The \trento{}~3D calculation returns a continuous density profile, and does not include hadronization. Therefore, the fluctuations calculated in this model are dynamical, and the subtraction of Poisson fluctuations explained in Sec.~\ref{s:subtraction} does not apply.
}
\end{table}

We now discuss the fluctuations of $E_T$ and $N_{\rm ch}$ for central collisions, and their mutual correlation. 
We use as a measure of these fluctuations the relative covariance matrix~\cite{Pruneau:2002yf}, defined as 
\begin{equation}
\label{defsigmaij}
\sigma_{ij}\equiv\frac{\langle N_iN_j\rangle}{\langle N_i\rangle\langle N_j\rangle}-1= \frac{\Sigma_{ij}}{\bar N_i\bar N_j}.
\end{equation}
Its diagonal elements are the relative variances of $N_i$. 
Values extracted from ATLAS data (Table~\ref{covariance}) show that the relative fluctuations of $E_T$ are smaller than those of $N_{\rm ch}$, both before and after isolating dynamical fluctuations. 
This is the reason why $E_T$ is a better estimator of the centrality than $N_{\rm ch}$~\cite{Aaboud:2019sma,Zhou:2018fxx}.
It is a non-trivial observation, which is not reproduced by the \trento{}~3D calculation. 
The  \trento{}~3D calculation reproduces the variance of $N_{\rm ch}$ to a good approximation, which is not surprising since it was fitted to the distribution of the charged multiplicity measured near mid-rapidity. 
On the other hand, it overestimates the variance of $E_T$ by at least a factor 3.\footnote{Note that the  \trento{}~3D  model only predicts the initial entropy density $s$. We have assumed $E_T\propto \int s^{4/3}$. If we had instead assumed $E_T\propto \int s$, the variance would be smaller by a factor $\sim (3/4)^2$. That is, the relative variance of $E_T$ would be $2.01\times 10^{-3}$ instead of $3.57\times 10^{-3}$, still larger than data by a factor $\sim 2$.}
It also overestimates the Pearson correlation coefficient between $E_T$ and $N_{\rm ch}$, defined as $\sigma_{12}/\sqrt{\sigma_{11}\sigma_{22}}$, whose value is $\sim 0.95$ in the \trento{}~3D calculation, and $\sim 0.85$ in data. 
In other words, the \trento{}~3D model largely underestimates the longitudinal decorrelation~\cite{Bozek:2017qir,Jia:2020tvb} between the rapidity windows where $E_T$ and $N_{\rm ch}$ are measured. 

There is no trivial relation between the relative fluctuations of $E_T$ and the relative fluctuations of the multiplicity in the same rapidity window.
In a hydrodynamic picture of the collision~\cite{Ollitrault:2007du}, however, one expects them to be of the same order of magnitude, because the multiplicity is proportional to the entropy, and the energy and the entropy are related through the equation of state.
In Sec.~\ref{s:conclusions}, we will suggest a specific analysis in order to check this assumption.
If it is correct, our results imply that multiplicity fluctuations are smaller at large rapidity than around central rapidity. 

The values of the relative variance of $E_T$ in Table~\ref{covariance} can be compared with those previously extracted from ATLAS data at $2.76$~TeV~\cite{Das:2017ned}. 
After subtracting Poisson fluctuations with help of multiplicity densities measured by ALICE~\cite{ALICE:2013jfw}, we obtain $\sigma_{11}=0.76\times 10^{-3}$. 
This is smaller than the value at $5.02$~TeV in Table~\ref{covariance}, which implies that multiplicity fluctuations increase as a function of the rapidity gap between incoming nuclei and the detector, $y_{\rm beam}-y$, which increases by $\sim 0.6$ between $2.76$~TeV and $5.02$~TeV. 
It is tempting to postulate that multiplicity fluctuations depend on $y_{\rm beam}-y$, much as average multiplicities themselves~\cite{PHOBOS:2005zhy,Gelis:2006tb}. 

\begin{figure}[ht]
\begin{center}
\includegraphics[width=\linewidth]{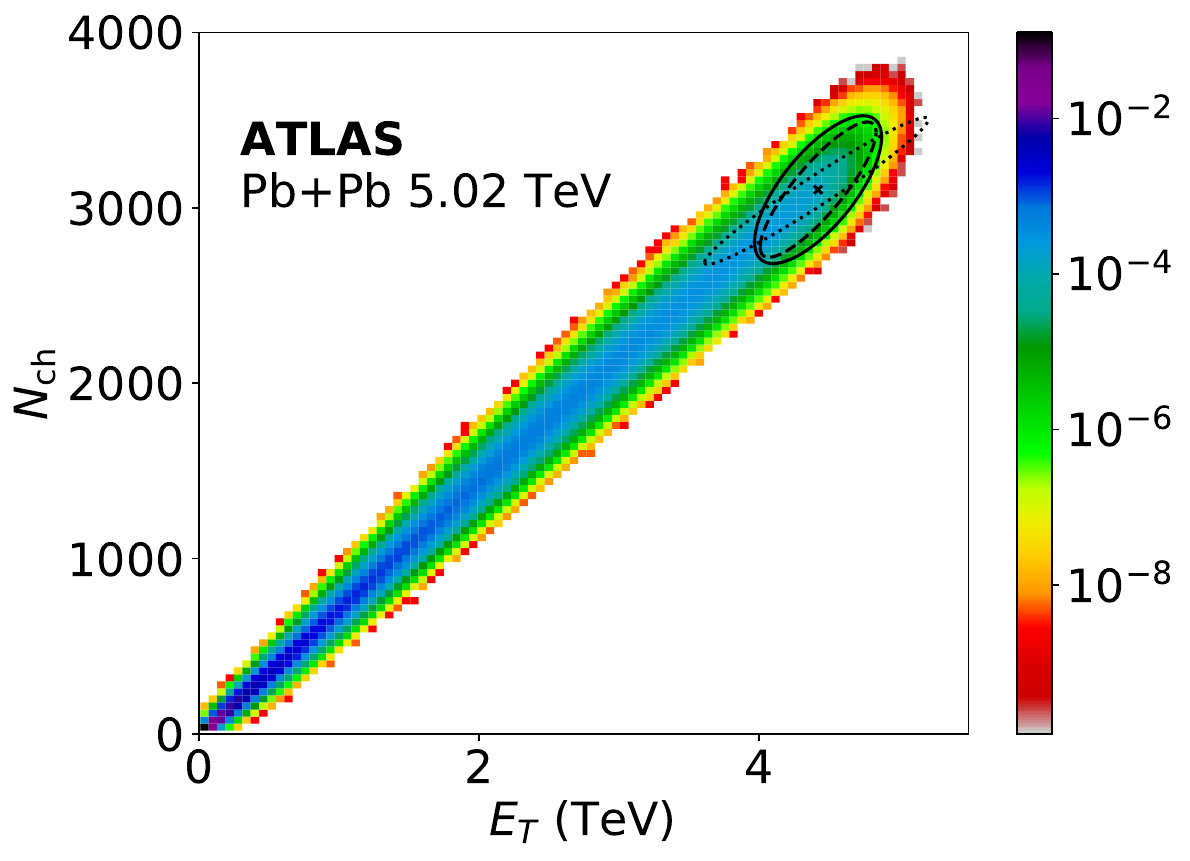} 
\end{center}
\caption{(Color online)
Full line:  $99\%$ confidence ellipse of zero impact parameter collisions, already shown in Fig.~\ref{fig:atlasillustration}. 
The ellipse is defined by Eq.~(\ref{defconfidenceellipse}).
The dashed line represents the ellipse, after subtraction of the contribution of Poisson fluctuations.
The dotted line is the value from the \trento{}~3D calculation. 
}
\label{fig:covellipse}
\end{figure}  

The second important observation is that $E_T$ and $N_{\rm ch}$ are strongly correlated, even at fixed impact parameter. 
This is illustrated by the elongated covariance ellipse represented in Fig.~\ref{fig:covellipse}.  
The correlation is significantly stronger after Poisson fluctuations have been subtracted out. 

We now interpret these results within a simple model. 
We assume that a nucleus-nucleus collision at impact parameter $b$ produces $N_s$ sources, and that $E_T$ and $N_{\rm ch}$ are obtained by summing the contributions of all sources in the event~\cite{Bozek:2017qir,Zhou:2018fxx,Jia:2020tvb}. 
We assume that the sources are independent.\footnote{ 
Note that a similar picture underlies our assumption of Gaussian fluctuations, as explained in Sec.~\ref{s:method}.}
We denote the contribution of a single source to $N_1=E_T$ and $N_2=N_{\rm ch}$ by $n_1$ and $n_2$, respectively. 
Under these assumptions, one easily obtains: 
\begin{equation}
  \label{modeleqs}
\sigma_{ij}=\frac{\langle N_s^2\rangle-\langle N_s\rangle^2}{\langle N_s\rangle^2}+
\frac{1}{\langle N_s\rangle}\frac{\langle n_i n_j\rangle-\langle n_i \rangle\langle n_j\rangle}{\langle n_i \rangle\langle n_j\rangle}
\end{equation}
The first term in the right-hand side is the relative variance of the number of sources, which gives the same contribution to all elements of the covariance matrix.  
It is a trivial correlation arising from a global change in the system, much as  correlation stemming from the variation of impact parameter~\cite{Kovchegov:2012nd}. 
It does not shed light on rapidity correlations in the emission process.
The last term in the right-hand side, which involves the covariance matrix of a single source, is physically more interesting. 

It is not possible to uniquely determine the quantities in the right-hand side of Eq.~(\ref{modeleqs}), because there are more unknowns than equations. 
But one can obtain non-trivial information. 
For instance, we have noted that the relative variance of $E_T$ is smaller than that of $N_{\rm ch}$. 
Since they get the same contribution from the variance of the number of sources, it implies that the difference between the relative variances of $E_T$  and $N_{\rm ch}$ is larger at the level of a single source.
The suppression of fluctuations at large rapidity is by no means a small effect. 
One can also obtain non-trivial information on the correlation between $n_1$ and $n_2$, which represent the contributions of a single source to  $E_T$ and $N_{\rm ch}$.
For instance, one can readily exclude that they are uncorrelated, because this would imply $\sigma_{11}>\sigma_{12}$, at variance with the values in Table~\ref{covariance}, after Poisson fluctuations have been subtracted out. 
One can obtain a lower bound $r_{\rm min}$ on the Pearson correlation coefficient $r$ between $n_1$ and $n_2$, defined by
\begin{equation}
r\equiv \frac{\langle n_1n_2\rangle-\langle n_1\rangle\langle n_2\rangle}{\sqrt{\langle n_1^2\rangle- \langle n_1\rangle^2}\sqrt{\langle n_2^2\rangle- \langle n_2\rangle^2}}.
\end{equation}
Elementary algebra shows that 
\begin{equation}
\label{perfectsquare}
r^2=r_{\rm min}^2+\left(\frac{r(c_v(n_1)^2+c_v(n_2)^2)-2c_v(n_1)c_v(n_2)}{c_v(n_1)^2-c_v(n_2)^2} \right)^2,
\end{equation}
where $c_v(n_i)\equiv \sqrt{\langle n_i^2\rangle/\langle n_i\rangle^2-1}$
is the coefficient of variation  (or relative standard deviation) of $n_i$, and 
\begin{equation}
r_{\rm min}\equiv 
2\frac{\sqrt{(\sigma_{12}-\sigma_{11})(\sigma_{22}-\sigma_{12})}}{\sigma_{22}-\sigma_{11}}.
\end{equation}
Equation~(\ref{perfectsquare}) guarantees that  $r\ge r_{\rm min}$, 
With the values in Table~\ref{covariance}, and taking into account the error bars (from the reconstruction and from the subtraction of Poisson fluctuations) we obtain $r_{\rm min}=0.72\pm 0.15$. 
Two effects may contribute to this strong correlation: 
Some sources may be stronger than others, for instance those situated in the center of the interaction region. 
Stronger sources yield larger values of both $n_1$ and $n_2$, which induces a mutual correlation. 
The second effect is the dynamical effect that one would like to isolate, that particle production from a single source is strongly correlated across rapidities. 
More detailed modeling will be necessary to disentangle the relative contributions of these two effects. 

\begin{figure}[ht]
\begin{center}
\includegraphics[width=\linewidth]{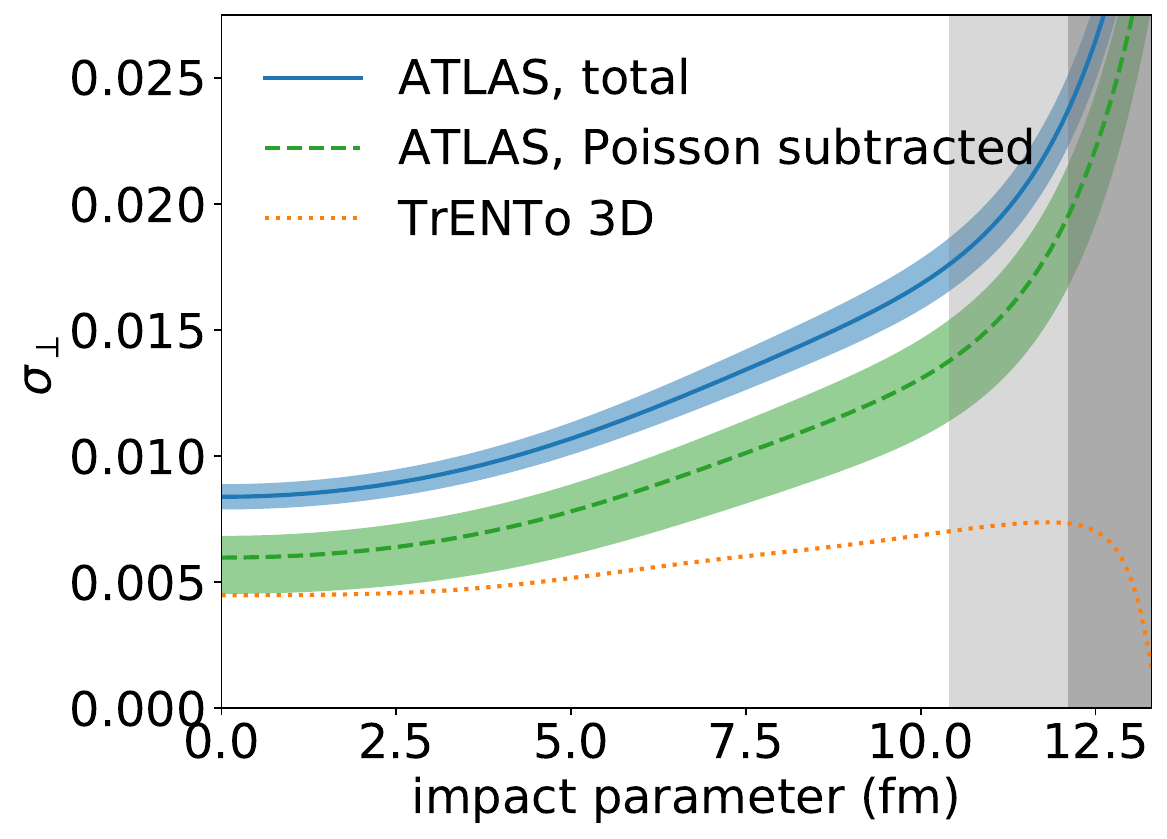} 
\end{center}
\caption{(Color online)
\label{fig:sigmaperpatlas}
 Same as Fig.~\ref{fig:nchatlas} for the width of the distribution, defined by Eq.~(\ref{defsigmaperp}).  
Full line: before subtracting Poisson fluctuations. 
  Dashed line: after subtracting Poisson fluctuations according to Eqs.~(\ref{etsubtraction}) and (\ref{nchsubtraction}). 
  The shaded area around the curve is our estimate of the error on the reconstruction. 
  Dotted line: value calculated in the \trento{}~3D model.
}
\end{figure}

\subsection{Impact parameter dependence of fluctuations}
\label{s:sigmaperp}

As explained in Sec.~\ref{s:method}, one cannot reconstruct the impact parameter dependence of the full covariance matrix. 
Only the specific linear combination $\sigma_\perp$, defined by Eq.~(\ref{defsigmaperp}), can be reconstructed for all impact parameters. 
Its impact parameter dependence is represented in Fig.~\ref{fig:sigmaperpatlas}, before and after subtracting  Poisson fluctuations (Sec.~\ref{s:subtraction}).  
The \trento{}~3D model underpredicts $\sigma_\perp$ for all values of impact parameter, which is the reason why the distribution in Fig.~\ref{fig:histogramtrento3d} is narrower than that in Fig.~\ref{fig:atlasillustration}. 
This shows that the width of the distribution, as measured by $\sigma_\perp$, contains non-trivial dynamical information, which can be used to discriminate between models.  

\section{Conclusions and perspectives}
\label{s:conclusions}

We have introduced a simple Bayesian method which allows for a robust reconstruction of multiplicity fluctuations and rapidity correlations in nucleus-nucleus collisions at fixed impact parameter $b$. 
Starting from the distribution $p(N_1,\cdots,N_p)$ of multiplicities (or transverse energies, or number of hits)  $N_1,\cdots,N_p$ measured in different parts of the detector, one can reconstruct the full covariance matrix of $(N_1,\cdots,N_p)$ at $b=0$, as well as the impact parameter dependence of a $(p-1)\times (p-1)$ projection of the matrix. 
We have applied the method to ATLAS data on the joint distribution of transverse energy and charged multiplicity.  
We have shown that dynamical fluctuations are smaller at large rapidity than around central rapidity, and that particle production is strongly correlated across rapidity, even at fixed impact parameter. 
We have also shown that the width of the distribution and its impact parameter dependence are not reproduced by our model calculation, so that this observable can be used to rule out three-dimensional models of initial fluctuations. 

Rapidity correlations have so far been studied using multiplicities of charged particle tracks~\cite{STAR:2009goo,ATLAS:2016rbh}.
However, particle tracks are typically reconstructed only around central rapidity, so that the rapidity coverage of such studies is limited. 
We have shown that track reconstruction is not needed. 
The transverse energy in a calorimeter, the multiplicity of pixel clusters~\cite{CMS:2013bza} or the number of hits in a scintillator~\cite{ALICE:2013hur} also give a quantitative information on fluctuations, provided that the detector is large enough that dynamical fluctuations dominate over Poisson fluctuations. 
A few tests should however be made in order to interpret correctly the information from calorimeters. 
Specifically,  one should first figure out how energy fluctuations relate to multiplicity fluctuations in the same rapidity window. 
This can be done easily using one of the central detectors of ALICE, CMS or ATLAS. 
These detectors measure the multiplicity of charged particle tracks and their momenta. 
One can use the sum of transverse momenta of charged particles, $P_T\equiv\sum_{N_{\rm ch}} p_t$, as a proxy for their transverse energy. 
By comparing the tails of the distributions of $N_{\rm ch}$ and $P_T$, one can readily check how their relative variances compare for central collisions.

We conclude by listing analyses  that would shed additional light on long-range correlation. 
The transverse energy is usually defined as $E_T=E_F+E_B$, where $E_F$ and $E_B$ are transverse energies in forward and backward calorimeters~\cite{CMS:2013bza,ATLAS:2016rbh}. 
If measured, the distribution $p(E_F,E_B)$ would provide direct information on forward-backward rapidity correlations. 
This analysis would be even simpler than that carried out in this paper because backward and forward calorimeters are symmetric around mid-rapidity, and this symmetry reduces the number of fit parameters in the Bayesian analysis. 
Similarly, the ALICE collaboration could measure the distribution $p(V_{0A},V_{0C})$, where  $V_{0A}$ and $V_{0C}$ are the multiplicities in backward and forward scintillators~\cite{Abelev:2013qoq}. 
These analyses can be readily carried out by experimental collaborations using existing data, by following the exact same steps as our analysis of ATLAS public data. 

More detailed information on the rapidity structure could be obtained by extending the analysis to three variables. 
For instance, in the case of the ATLAS data studied in this paper~\cite{Aaboud:2019sma}, one could split the calorimeter into its forward and backward components and measure $p(E_F,E_B,N_{\rm ch})$. 
Similar analyses could be done with the CMS and ALICE detectors. 
They would yield detailed information on the long-range rapidity structure of correlations. 
Finally, the same method could be applied to heavy-ion collisions at lower energies~\cite{INDRA:2020kyj,Parfenov:2021ipw,Li:2021plq}.

\section*{Acknowledgements}
We thank J.~Jia for sending us the centrality calibration of data.
AK thanks Bhavya Bhatt for technical assistance in
the early stages of this work. 
JYO thanks F.~Gelis and G.~Giacalone for discussions. 
RSB would like to thank the Department of Science and Technology,
India for grant no. SERB /CRG/2019/000807.

\appendix
\section{Reconstruction of the covariance matrix in non-central collisions}
\label{s:saddlepoint}

We have shown in Secs.~\ref{s:validation} and \ref{s:data} that for $b=0$, the whole $(p\times p)$ covariance matrix $\Sigma_{ij}(b=0)$ can be accurately reconstructed from data. 
In this Appendix, we specify which information about $\Sigma_{ij}(b)$ can be reconstructed for $b>0$. 
Inserting Eq.~(\ref{expansion}) into Eq.~(\ref{nfixedb}) and neglecting the pre-exponential factor, one obtains:  
\begin{equation}
\label{nfixedb2}
P(N_1,...,N_p|b)\propto
\exp\left(-\frac{1}{2}\sum\limits_{ij} (c_i-c_b)\bar N'_i\Sigma^{-1}_{ij}\bar N'_j(c_j-c_b)\right),  
\end{equation}
where we omit the dependence of $\bar N'_i$ and $\Sigma^{-1}_{ij}$ on $b$ for simplicity, and $c_i$ is a shorthand for $c_i(N_i)$. 
Thus the integral in Eq.~(\ref{cbint}) is a Gaussian integral over $c_b$. 
For a point in the bulk of the distribution (see Fig.~\ref{fig:illusdeltac}), the maximum of the integrand is at $c_b>0$. 
The exponential decays very fast away from the maximum, so that the integral over $c_b$ can be evaluated from $-\infty$ to $+\infty$.\footnote{By contrast, for a point in the tip of the distribution, corresponding to a very central collision, the maximum of the integrand is at $c_b=0$, and only values of $c_b$ very close to $0$ contribute significantly.} 
Neglecting again the pre-exponential factor, one obtains
\begin{equation}
\label{saddlepoint2}
P(N_1,...,N_p)\propto
\exp\left(-\frac{1}{2}\sum_{ij} c_i\Pi_{ij}c_j\right), 
\end{equation}
where
\begin{equation}
\label{defpiij}
\Pi_{ij}\equiv \bar N'_i\Sigma^{-1}_{ij}\bar N'_j-\frac{\left(\sum_{\alpha}  \bar N'_i\Sigma^{-1}_{i\alpha}\bar N'_\alpha\right)\left(\sum_{\beta} \bar N'_j\Sigma^{-1}_{j\beta}\bar N'_\beta\right)}{\sum\limits_{\alpha\beta}\bar N'_\alpha\Sigma^{-1}_{\alpha\beta}\bar N'_\beta}.
\end{equation}
$\mathbf{\Pi}$ is the quantity which can be reconstructed for all impact parameters. 
It is a $(p\times p)$ symmetric matrix which verifies the property $\sum_i \Pi_{ij}=0$ for all $j$, that is, all lines and columns sum up to zero. 
Due to this property, one can rewrite Eq.~(\ref{saddlepoint2}) as:
\begin{equation}
\label{saddlepoint3}
P(N_1,...,N_p)\propto
\exp\left(-\frac{1}{2}\sum_{i>1,j>1} (c_i-c_1)\Pi_{ij}(c_j-c_1)\right).
\end{equation}
This form shows that the distribution of $(N_1,...,N_p)$ is a function of $p-1$ variables $c_i-c_1$. 
Physically, $\mathbf{\Pi}$ represents the projection of the covariance matrix onto the $(p-1)$-dimensional subspace orthogonal to the ridge line. 
In the case $p=2$, Eq.~(\ref{defpiij}) gives
\begin{equation}
\label{piijp=2}
\mathbf{\Pi}=\frac{1}{\sigma_\perp^2}
\begin{pmatrix}\phantom{-}1&-1\\ -1&\phantom{-}1\end{pmatrix},
\end{equation}
where $\sigma_\perp$ is defined by Eq.~(\ref{defsigmaperp}). 
Inserting Eq.~(\ref{piijp=2}) into Eq.~(\ref{saddlepoint3}), one recovers Eq.~(\ref{result2d}).


\begin{thebibliography}{99}
\bibitem{Lappi:2006fp}
T.~Lappi and L.~McLerran,
Nucl. Phys. A \textbf{772}, 200-212 (2006)
doi:10.1016/j.nuclphysa.2006.04.001
[arXiv:hep-ph/0602189 [hep-ph]].

\bibitem{Andersson:1983ia}
B.~Andersson, G.~Gustafson, G.~Ingelman and T.~Sjostrand,
Phys. Rept. \textbf{97}, 31-145 (1983)
doi:10.1016/0370-1573(83)90080-7

\bibitem{Gelis:2008sz}
F.~Gelis, T.~Lappi and R.~Venugopalan,
Phys. Rev. D \textbf{79}, 094017 (2009)
doi:10.1103/PhysRevD.79.094017
[arXiv:0810.4829 [hep-ph]].

\bibitem{Lappi:2019kif}
T.~Lappi and A.~Ramnath,
Phys. Rev. D \textbf{100}, no.5, 054003 (2019)
doi:10.1103/PhysRevD.100.054003
[arXiv:1904.00782 [hep-ph]].

\bibitem{Bzdak:2012tp}
A.~Bzdak and D.~Teaney,
Phys. Rev. C \textbf{87}, no.2, 024906 (2013)
doi:10.1103/PhysRevC.87.024906
[arXiv:1210.1965 [nucl-th]].

\bibitem{Olszewski:2017vyg}
A.~Olszewski and W.~Broniowski,
Phys. Rev. C \textbf{96}, no.5, 054903 (2017)
doi:10.1103/PhysRevC.96.054903
[arXiv:1706.02862 [nucl-th]].

\bibitem{PHOBOS:2006mfc}
B.~B.~Back \textit{et al.} [PHOBOS],
Phys. Rev. C \textbf{74}, 011901 (2006)
doi:10.1103/PhysRevC.74.011901
[arXiv:nucl-ex/0603026 [nucl-ex]].

\bibitem{STAR:2009goo}
B.~I.~Abelev \textit{et al.} [STAR],
Phys. Rev. Lett. \textbf{103}, 172301 (2009)
doi:10.1103/PhysRevLett.103.172301
[arXiv:0905.0237 [nucl-ex]].

\bibitem{ATLAS:2016rbh}
M.~Aaboud \textit{et al.} [ATLAS],
Phys. Rev. C \textbf{95}, no.6, 064914 (2017)
doi:10.1103/PhysRevC.95.064914
[arXiv:1606.08170 [hep-ex]].

\bibitem{ALICE:2017mtc}
S.~Acharya \textit{et al.} [ALICE],
Phys. Lett. B \textbf{781}, 20-32 (2018)
doi:10.1016/j.physletb.2018.03.051
[arXiv:1710.07975 [nucl-ex]].

\bibitem{Aaboud:2019sma}
M.~Aaboud \textit{et al.} [ATLAS],
JHEP \textbf{01}, 051 (2020)
doi:10.1007/JHEP01(2020)051
[arXiv:1904.04808 [nucl-ex]].

\bibitem{Bernhard:2016tnd}
J.~E.~Bernhard, J.~S.~Moreland, S.~A.~Bass, J.~Liu and U.~Heinz,
Phys. Rev. C \textbf{94}, no.2, 024907 (2016)
doi:10.1103/PhysRevC.94.024907
[arXiv:1605.03954 [nucl-th]].

\bibitem{Nijs:2020roc}
G.~Nijs, W.~van der Schee, U.~G\"ursoy and R.~Snellings,
Phys. Rev. C \textbf{103}, no.5, 054909 (2021)
doi:10.1103/PhysRevC.103.054909
[arXiv:2010.15134 [nucl-th]].

\bibitem{JETSCAPE:2020mzn}
D.~Everett \textit{et al.} [JETSCAPE],
Phys. Rev. C \textbf{103}, no.5, 054904 (2021)
doi:10.1103/PhysRevC.103.054904
[arXiv:2011.01430 [hep-ph]].

\bibitem{Parkkila:2021tqq}
J.~E.~Parkkila, A.~Onnerstad and D.~J.~Kim,
Phys. Rev. C \textbf{104}, no.5, 054904 (2021)
doi:10.1103/PhysRevC.104.054904
[arXiv:2106.05019 [hep-ph]].

\bibitem{Ke:2016jrd}
W.~Ke, J.~S.~Moreland, J.~E.~Bernhard and S.~A.~Bass,
Phys. Rev. C \textbf{96}, no.4, 044912 (2017)
doi:10.1103/PhysRevC.96.044912
[arXiv:1610.08490 [nucl-th]].

\bibitem{Miller:2007ri}
M.~L.~Miller, K.~Reygers, S.~J.~Sanders and P.~Steinberg,
Ann. Rev. Nucl. Part. Sci. \textbf{57}, 205-243 (2007)
doi:10.1146/annurev.nucl.57.090506.123020
[arXiv:nucl-ex/0701025 [nucl-ex]].

\bibitem{dEnterria:2020dwq}
D.~d'Enterria and C.~Loizides,
Ann. Rev. Nucl. Part. Sci. \textbf{71}, 315-44 (2021)
doi:10.1146/annurev-nucl-102419-060007
[arXiv:2011.14909 [hep-ph]].

\bibitem{Das:2017ned}
S.~J.~Das, G.~Giacalone, P.~A.~Monard and J.~Y.~Ollitrault,
Phys. Rev. C \textbf{97}, no.1, 014905 (2018)
doi:10.1103/PhysRevC.97.014905
[arXiv:1708.00081 [nucl-th]].

\bibitem{Busza:2018rrf}
W.~Busza, K.~Rajagopal and W.~van der Schee,
Ann. Rev. Nucl. Part. Sci. \textbf{68}, 339-376 (2018)
doi:10.1146/annurev-nucl-101917-020852
[arXiv:1802.04801 [hep-ph]].

\bibitem{Broniowski:2007nz}
W.~Broniowski, M.~Rybczynski and P.~Bozek,
Comput. Phys. Commun. \textbf{180}, 69-83 (2009)
doi:10.1016/j.cpc.2008.07.016
[arXiv:0710.5731 [nucl-th]].

\bibitem{Moreland:2014oya}
J.~S.~Moreland, J.~E.~Bernhard and S.~A.~Bass,
Phys. Rev. C \textbf{92}, no.1, 011901 (2015)
doi:10.1103/PhysRevC.92.011901
[arXiv:1412.4708 [nucl-th]].

\bibitem{Loizides:2017ack}
C.~Loizides, J.~Kamin and D.~d'Enterria,
Phys. Rev. C \textbf{97}, no.5, 054910 (2018)
[erratum: Phys. Rev. C \textbf{99}, no.1, 019901 (2019)]
doi:10.1103/PhysRevC.97.054910
[arXiv:1710.07098 [nucl-ex]].

\bibitem{Bierlich:2018xfw}
C.~Bierlich, G.~Gustafson, L.~L\"onnblad and H.~Shah,
JHEP \textbf{10}, 134 (2018)
doi:10.1007/JHEP10(2018)134
[arXiv:1806.10820 [hep-ph]].

\bibitem{Bozek:2019wyr}
P.~Bo\.zek, W.~Broniowski, M.~Rybczynski and G.~Stefanek,
Comput. Phys. Commun. \textbf{245}, 106850 (2019)
doi:10.1016/j.cpc.2019.07.014
[arXiv:1901.04484 [nucl-th]].

\bibitem{Abelev:2013qoq}
B.~Abelev \textit{et al.} [ALICE],
Phys. Rev. C \textbf{88}, no.4, 044909 (2013)
doi:10.1103/PhysRevC.88.044909
[arXiv:1301.4361 [nucl-ex]].

\bibitem{Broniowski:2001ei}
W.~Broniowski and W.~Florkowski,
Phys. Rev. C \textbf{65}, 024905 (2002)
doi:10.1103/PhysRevC.65.024905
[arXiv:nucl-th/0110020 [nucl-th]].

\bibitem{Rogly:2018ddx}
R.~Rogly, G.~Giacalone and J.~Y.~Ollitrault,
Phys. Rev. C \textbf{98}, no.2, 024902 (2018)
doi:10.1103/PhysRevC.98.024902
[arXiv:1804.03031 [nucl-th]].

\bibitem{ALICE:2015juo}
J.~Adam \textit{et al.} [ALICE],
Phys. Rev. Lett. \textbf{116}, no.22, 222302 (2016)
doi:10.1103/PhysRevLett.116.222302
[arXiv:1512.06104 [nucl-ex]].

\bibitem{ALICE:2016fbt}
J.~Adam \textit{et al.} [ALICE],
Phys. Lett. B \textbf{772}, 567-577 (2017)
doi:10.1016/j.physletb.2017.07.017
[arXiv:1612.08966 [nucl-ex]].

\bibitem{ALICE:2018hza}
S.~Acharya \textit{et al.} [ALICE],
Phys. Lett. B \textbf{788}, 166-179 (2019)
doi:10.1016/j.physletb.2018.10.052
[arXiv:1805.04399 [nucl-ex]].

\bibitem{DiFrancesco:2016srj}
P.~Di Francesco, M.~Guilbaud, M.~Luzum and J.~Y.~Ollitrault,
Phys. Rev. C \textbf{95}, no.4, 044911 (2017)
doi:10.1103/PhysRevC.95.044911
[arXiv:1612.05634 [nucl-th]].

\bibitem{Mazeliauskas:2018irt}
A.~Mazeliauskas, S.~Floerchinger, E.~Grossi and D.~Teaney,
Eur. Phys. J. C \textbf{79}, no.3, 284 (2019)
doi:10.1140/epjc/s10052-019-6791-7
[arXiv:1809.11049 [nucl-th]].

\bibitem{ALICE:2013mez}
B.~Abelev \textit{et al.} [ALICE],
Phys. Rev. C \textbf{88}, 044910 (2013)
doi:10.1103/PhysRevC.88.044910
[arXiv:1303.0737 [hep-ex]].

\bibitem{Pruneau:2002yf}
C.~Pruneau, S.~Gavin and S.~Voloshin,
Phys. Rev. C \textbf{66}, 044904 (2002)
doi:10.1103/PhysRevC.66.044904
[arXiv:nucl-ex/0204011 [nucl-ex]].

\bibitem{Zhou:2018fxx}
M.~Zhou and J.~Jia,
Phys. Rev. C \textbf{98}, no.4, 044903 (2018)
doi:10.1103/PhysRevC.98.044903
[arXiv:1803.01812 [nucl-th]].

\bibitem{Bozek:2017qir}
P.~Bozek and W.~Broniowski,
Phys. Rev. C \textbf{97}, no.3, 034913 (2018)
doi:10.1103/PhysRevC.97.034913
[arXiv:1711.03325 [nucl-th]].

\bibitem{Jia:2020tvb}
J.~Jia, C.~Zhang and J.~Xu,
Phys. Rev. Res. \textbf{2}, no.2, 023319 (2020)
doi:10.1103/PhysRevResearch.2.023319
[arXiv:2001.08602 [nucl-th]].

\bibitem{Ollitrault:2007du}
J.~Y.~Ollitrault,
Eur. J. Phys. \textbf{29}, 275-302 (2008)
doi:10.1088/0143-0807/29/2/010
[arXiv:0708.2433 [nucl-th]].

\bibitem{ALICE:2013jfw}
E.~Abbas \textit{et al.} [ALICE],
Phys. Lett. B \textbf{726}, 610-622 (2013)
doi:10.1016/j.physletb.2013.09.022
[arXiv:1304.0347 [nucl-ex]].

\bibitem{PHOBOS:2005zhy}
B.~B.~Back \textit{et al.} [PHOBOS],
Phys. Rev. C \textbf{74}, 021901 (2006)
doi:10.1103/PhysRevC.74.021901
[arXiv:nucl-ex/0509034 [nucl-ex]].

\bibitem{Gelis:2006tb}
F.~Gelis, A.~M.~Stasto and R.~Venugopalan,
Eur. Phys. J. C \textbf{48}, 489-500 (2006)
doi:10.1140/epjc/s10052-006-0020-x
[arXiv:hep-ph/0605087 [hep-ph]].

\bibitem{Kovchegov:2012nd}
Y.~V.~Kovchegov and D.~E.~Wertepny,
Nucl. Phys. A \textbf{906}, 50-83 (2013)
doi:10.1016/j.nuclphysa.2013.03.006
[arXiv:1212.1195 [hep-ph]].

\bibitem{CMS:2013bza}
S.~Chatrchyan \textit{et al.} [CMS],
JHEP \textbf{02}, 088 (2014)
doi:10.1007/JHEP02(2014)088
[arXiv:1312.1845 [nucl-ex]].

\bibitem{ALICE:2013hur}
B.~Abelev \textit{et al.} [ALICE],
Phys. Rev. C \textbf{88}, no.4, 044909 (2013)
doi:10.1103/PhysRevC.88.044909
[arXiv:1301.4361 [nucl-ex]].

\bibitem{INDRA:2020kyj}
J.~D.~Frankland \textit{et al.} [INDRA],
Phys. Rev. C \textbf{104}, no.3, 034609 (2021)
doi:10.1103/PhysRevC.104.034609
[arXiv:2011.04496 [nucl-ex]].

\bibitem{Parfenov:2021ipw}
P.~Parfenov, D.~Idrisov, V.~B.~Luong and A.~Taranenko,
Particles \textbf{4}, no.2, 275-287 (2021)
doi:10.3390/particles4020024

\bibitem{Li:2021plq}
F.~Li, Y.~Wang, Z.~Gao, P.~Li, H.~L\"u, H.~Lv, Q.~Li, C.~Y.~Tsang and M.~B.~Tsang,
Phys. Rev. C \textbf{104}, no.3, 034608 (2021)
doi:10.1103/PhysRevC.104.034608
[arXiv:2105.08912 [nucl-th]].
\end{thebibliography}
\end{document}